\documentclass[a4paper,10pt]{article}

\usepackage{url}
\usepackage{amsfonts}
\usepackage{amsmath}
\usepackage{amssymb}
\usepackage{mathrsfs}
\usepackage{subfigure}
\usepackage{graphicx}
\usepackage{epstopdf}
\usepackage{float}
\usepackage{color}
\usepackage{bm}
\usepackage{ulem}
\usepackage{xcolor}
\usepackage{appendix}
\usepackage{booktabs}
\usepackage{hyperref}
\usepackage{array}
\usepackage{cancel}
\usepackage{footnote}
\usepackage{cite}
\usepackage{authblk}  
\usepackage{geometry} 
 \allowdisplaybreaks[2]

\begin{document}

\title{Gravitational Lensing of Euler-Heisenberg Black Hole Surrounded by Perfect Fluid Dark Matter}

\author{Ping Su \thanks{Email: suping@cqut.edu.cn} and Chen-Kai Qiao \thanks{Email: chenkaiqiao@cqut.edu.cn}}

\affil{College of Science, Chongqing University of Technology, Banan, Chongqing, 400054, China}

\maketitle
	
\begin{abstract}
In this work, we study the gravitational lensing of Euler-Heisenberg black hole surrounded by perfect fluid dark matter. This kind of black hole solution enables us to investigate the nontrivial interplay between the dark matter effects and nonlinear electrodynamics effects (or quantum electrodynamics effects) on charged black hole systems. The important observables in gravitational lensings are calculated and discussed in our work, including the gravitational deflection angle of light and time delay of light. Additionally, we also explore the massive orbit's bound orbits (and their precession angles) and black hole shadow radius for Euler-Heisenberg black hole in the presence of dark matter. The results indicate that the Euler-Heisenberg black hole with a larger perfect fluid dark matter parameter could greatly reduce the gravitational deflection angle of light, time delay of light and precession angle of massive object's bound orbit, while the nonlinear electrodynamics effects do not have large influences on these observables.

\ \ \ 


\ \ \ \ \ \ \ \ \ \ \ \ \ \ \ \ \ 
Perfect Fluid Dark Matter; Nonlinear Electrodynamics Effects 
\end{abstract}

\section{Introduction}

In the past few decades, modern astrophysical observations reveal that most of our universe is dominated by dark matter and dark energy. Particularly, the cosmic microwave background (CMB) observations suggested that $26.8\%$ of our universe is consist of dark matter, $68.3\%$ of our universe is composed of dark energy, while only $4.9\%$ of our universe is made up of baryonic matter \cite{Planck2013,Planck2013b,Planck2018,Planck2018b}. Theoretically, the dark matter is composed of unknown particles predicted in theories beyond the Standard Model, such as axions, sterile neutrinos and Weakly Interacting Massive Particles (WIMPs). In astronomy, the dark matter has extremely strong influences on the structure and formation of galaxy clusters, and it can also dominantly affect the star motions in galaxies and galaxy clusters, producing observational flat rotational curves in spiral galaxies \cite{Rubin1970,Rubin2001,Corbelli2000,Bertone2018}. Notably, recent studies also suggested that galactic dark matter halos have non-negligible influences on the gravitational lensing and shadow images of supermassive black holes \cite{XuZ2018,HouX2018,Jusufi2019a,Jusufi2019b,Jusufi2020,Jusufi2021,Pantig2021,Pantig2022a,Pantig2022b,Pantig2023,Konoplya2022,Xavier2023,LiuD2023a,LiuD2023b,LiuD2024,QiaoCK2023a,QiaoCK2024,Pantig2024}.

The equation of state for dark matter in galaxies (around the supermassive black hole in the galaxy center) is still a mystery. Recently, a perfect fluid dark matter (PFDM) model has attracted great interest \cite{Rahaman2011,Barranco2015,Potapov2016}. Despite the simple form of PFDM model, it can reflect some non-trivial effects and properties of dark matter. In recent years, a number of studies investigate the PFDM effects on many aspects of black hole physics, and effective spacetime metrics for several kinds of black holes (such as Schwarzschild, Reissner-Nordstr\"om, Kerr black holes) surrounded by PFDM have been successfully constructed \cite{Kiselev2003a,Kiselev2003b,LiMH2012,Heydarzade2017,Das2021}. Based on these effective spacetime metrics, the gravitational lensing, black hole shadow images, quasi-normal mode, thermodynamic properties of black hole system in PFDM medium can be extensively analyzed \cite{XuZ2018a,XuZ2018b,XuZ2019,Rizwan2018,Haroon2018,HouX2018b,XuZ2016,Narzilloev2020,Atamurotov2021,TaoJ2021,TaoJ2023,HuYP2023,Fard2023,Abbas2023,QiaoCK2023,Das2022,Das2024,YangXT2024,Ditta2024}. 

Inspired by the simplicity and success of PFDM model in black hole physics and astronomy, we choose a novel black hole solution in PFDM medium --- the Euler-Heisenberg black hole surrounded by PFDM. The Euler-Heisenberg black hole, which is obtained from the Euler-Heisenberg effective action \cite{Yajima2001,Salazar1987,Kruglov2017,Breton2019,Nashed2021}, stimulated numerous studies in past few years, for its nontrivial realization on the coupling between nonlinear electrodynamics and gravity systems, especially with the quantum loop-corrections in quantum electrodynamics (QED) \cite{Heisenberg1936}. With these desired properties, this kind of black hole is often used to explore the nonlinear electrodynamics in black hole physics, evoking a number of studies on thermodynamic transitions, thermodynamic topology, black hole shadow, gravitational lensing, quasi-normal modes for black hole systems with nonlinear electrodynamics in recent years \cite{Breton2016,Gulin2017,Allahyari2019,Magos2020,Olvera2020,Amaro2020,Maceda2020,GuoMY2021a,GuoMY2021b,LiuYX2021,HuYP2021,Breton2021,Amaro2022,Ovgun2022,Javed2022,Nomura2022,ZengXX2022,LuoZ2022,YeX2022,DaiH2022,LiGR2022,ChenDY2022,Alipour2023,WangBQ2023,YuQ2023,Breton2022,Rehman2023,Ali2024,Gogoi2024,Theodosopoulos2024,Sekhmani2024,Bolokhov2024,Lambiase2024,Soares2023,Ditta2023c}. The nonlinear electrodynamics effects (or QED effects) on gravitation are extremely important issues in high energy scale, and they may have potential influences on the evolution of our universe (especially at the beginning stage), such as the inflationary theory in Big Bang, the dynamic evolution of dark matter and dark energy, and the cosmological accelerating expansion of universe. Therefore, it is interesting to study the interplay between dark matter and nonlinear electrodynamics in gravitational systems. The Euler-Heisenberg black hole surrounded by PFDM could provide us a simple way to investigate its influences on charged black holes. Recently, several analytical solutions for Euler-Heisenberg black hole in the presence of dark matter medium have been constructed \cite{HuXR2024a,Yildiz2024,HuXR2024b,Hamil2024}, which enable us to give a comprehensive exploration on the interplay between dark matter effects and nonlinear electrodynamics effects in black hole systems.

In this work, we carry out an investigation on the gravitational lensing of Euler-Heisenberg black hole surrounded by PFDM. The gravitational lensing, which is caused by spacetime curvature and could result in distortion or deflection effects on photon propagation, offers us an avenue to reveal the features of gravitational field and intrinsic properties of black holes through high precision observational data \cite{Wambsganss1998}. The Einstein's general relativity, alternative gravity theories and astrophysical models can be tested from gravitational lensings \cite{Wambsganss1998,Mollerach2002,Schmidt2008}. In recent years, significant progress has been made on exploring gravitational lensing in large numbers of systems \cite{Virbhadra2000,Virbhadra2007,Gibbons2008,Chowdhuri2021,Chowdhuri2023,Ditta2023a,Ditta2023b,Soares2024b}. Particularly, the dark matter in galaxies and dark energy in our universe can be studied through gravitational lensing observations \cite{Clowe2006,Uitert2012,Brimioulle2013}. The amounts of dark matter and the properties of dark matter halos in galaxies and galaxies clusters can be strongly constrained by gravitational lensing observations \cite{Zumalacarregui2018}. Furthermore, recent developments showed that the electrodynamics and plasma medium in galaxies may greatly influence the gravitational lensing \cite{Bisnovatyi-Kogan2017,Crisnejo2018,Ali2024b,Ali2024c}. Therefore, an exploration of gravitational lensing observables for Euler-Heisenberg black hole surrounded by PFDM could enrich our knowledge on the gravitational field for charged black hole systems, especially with the influences coming from the dark matter effects and nonlinear electrodynamics effects (or QED effects). Concretely, in the present work, we calculate the gravitational deflection angle of light and time delay of light for Euler-Heisenberg black hole surrounded by PFDM. They are significantly important observables in gravitational lensing, which have been frequently used to test various aspects of black hole physics and gravitational theories in both strong gravitational field limit and weak gravitational field limit. 

In addition to gravitational lensing observables, other observational quantities could also enrich our knowledge on black hole systems and gravitational fields. The precession of massive object's bound orbits near a massive celestial body is extremely important in understanding the central gravitational sources \cite{Levin2008,Chakraborty2014,Farrugia2016,El-Badry2023}. For example, the anomalous precession of Mercury's orbit is a renowned example in testing the Einstein's general relativity \cite{Will2014,Will2018}, and the motion of S stars around Sgr A* provides a pathway to study the property of supermassive black hole in our galaxy \cite{Eckart2002,GRAVITY2018,GRAVITY2020}. In our work, the massive object's bound orbits and their precession angles for Euler-Heisenberg black hole surrounded by PFDM are calculated and analyzed. 
Furthermore, the black hole shadow and black hole's optical images provide direct ways to probe the various effects (eg., strong gravitational field effects, electromagnetic effects, matter fields' effects) on black hole systems. A number of researches on black hole shadow have been stimulated since the capture of black hole images by Event Horizon Telescope (EHT) \cite{EHT2019a,EHT2019b,EHT2022,Perlick2022}. In the presented work, the black hole critical shadow radius for Euler-Heisenberg black hole in the presence of PFDM is also discussed.

The rest of the present work is organized in the following way. Section \ref{section2} briefly introduces the spacetime metric generated by Euler-Heisenberg black hole surrounded by PFDM. In section \ref{section3}, the theoretical approach employed in calculating the gravitational lensing observables is described in details. In section \ref{section4}, numerical results and discussions on the gravitational lensing observables (gravitational deflection angle of light and time delay of light), trajectory and precession angle of massive object's bound orbit, black hole shadow radius for Euler-Heisenberg black holes surrounded by PFDM are presented. Summary and perspectives are given in section \ref{section5}. Throughout this work, the natural unit $G=c=1$ is adopted.

\section{Euler-Heisenberg Black Hole Surrounded by Perfect Fluid Dark Matter} \label{section2}

The Euler-Heisenberg black hole becomes attractive in recent years, for its ability to include the nonlinear electrodynamics effects (or QED effects). This kind of black hole solution can be solved from the gravitational field equation and the Euler-Heisenberg effective action \cite{Yajima2001}, and it is one of the most important realization on the coupling between nonlinear electrodynamics and charged black hole systems.

Recently, several analytical solutions for Euler-Heisenberg black hole in the presence of dark matter medium have been constructed \cite{Yildiz2024,HuXR2024a,HuXR2024b,Hamil2024}. Particularly, S.-J. Ma \textit{et al.} derived the effective spacetime metric for an Euler-Heisenberg black hole surrounded by PFDM, through solving the gravitational field equation coupled with nonlinear electromagnetic field (with QED effects included) and dark matter field (assumed to be PFDM) \cite{HuXR2024a}. The action of a gravitational system coupled with nonlinear electromagnetic field and dark matter field can be written as \cite{HuXR2024a,HuXR2024b} 
\begin{equation}
	S = \frac{1}{4\pi} \int \sqrt{-g} \cdot d^{4}x 
	    \cdot \bigg[ \frac{R}{4} + \mathcal{L}_{\text{EM}} + \mathcal{L}_{\text{PFDM}} \bigg] ,
\end{equation}
and the nonlinear electromagnetic field's Lagrangian density is given by \cite{Heisenberg1936}
\begin{equation}
	\mathcal{L}_{\text{EM}} = - F + \frac{a}{2} F^2 + \frac{7a}{8} G^{2} ,
\end{equation}
where $F = \frac{1}{4} F_{\mu\nu} F^{\mu\nu}$ is the conventional electromagnetic invariant in Maxwell theory and $G = \frac{1}{4} F_{\mu\nu} *F^{\mu\nu}$ (the $*F_{\mu\nu}$ is the dual form of the electromagnetic strength tensor $F_{\mu\nu}$). The above Lagrangian leads to the gravitational field equation
\begin{equation}
	R_{\mu\nu} - \frac{1}{2} R g_{\mu\nu} = 8 \pi T_{\mu\nu} ,
\end{equation}
with the energy-momentum tensor contributed from nonlinear electromagnetic field and dark matter field. To derive a spherically symmetric analytical solution for black hole in this theory, the energy-momentum tensor of matter fields must be assumed to be a spherically symmetric distribution. In the work from S.-J. Ma \textit{et al.}, the energy-momentum tensor of matter fields has the following explicit form in spherical coordinates. The total energy-momentum tensor is composed of two parts: the energy-momentum of dark matter field and the energy-momentum of nonlinear electromagnetic field  \cite{HuXR2024a}
\begin{eqnarray}
	 T^{\mu}_{\nu}
	 & = & (T_{\text{EM}})_{\nu}^{\mu} + (T_{\text{DM}})_{\nu}^{\mu} \nonumber
	 \\
	 & = & \text{diag} \{ T^{t}_{t}, \ T^{r}_{r}, \ T^{\theta}_{\theta}, \ T^{\phi}_{\phi} \} , \nonumber
	 \\
	 (T_{\text{EM}})_{\nu}^{\mu} 
	 & = & \frac{1}{4\pi} \cdot 
	 \text{diag} \bigg\{ -\frac{Q^{2}}{2r^{4}}+\frac{aQ^{4}}{8r^{8}}, \  -\frac{Q^{2}}{2r^{4}}+\frac{aQ^{4}}{8r^{8}}, \  \frac{Q^{2}}{2r^{4}}-\frac{3aQ^{4}}{8r^{8}}, \  \frac{Q^{2}}{2r^{4}}-\frac{3aQ^{4}}{8r^{8}}  \bigg\} , \ \ \text{Electromagnetic Field} \ \ \ \  \nonumber
	 \\
	 (T_{\text{DM}})_{\nu}^{\mu} & = & 
	 \text{diag} \bigg\{ \frac{\lambda_{\text{DM}}}{8\pi r^{3}},\ \frac{\lambda_{\text{DM}}}{8\pi r^{3}}, \ -\frac{\lambda_{\text{DM}}}{16\pi r^{3}}, \ -\frac{\lambda_{\text{DM}}}{16\pi r^{3}} \bigg\} , \ \ \ \ \ \ \ \ \ \ \ \ \ \ \ \ \ \ \ \ \ \ \ \ \ \ \ \ \ \ \ \ \ \ \ \ \ \ \ \ \ \ \text{Dark Matter Field} 
	 \label{energy-momentum tensor}
\end{eqnarray}
where $M$ is the black hole mass, $Q$ is the black hole electric charge, $a$
is a parameter which measures the strength of nonlinear electrodynamics effects (or QED effects). The parameter $\lambda_{\text{DM}}$ denotes the influence from PFDM, which is proportional to the dark matter density. Eventually, by solving the gravitational field equation in the spherical coordinates, the spacetime metric for an Euler-Heisenberg black hole surrounded by PFDM is
\begin{equation}
	ds^2 = -f(r) dt^2 + \frac{1}{f(r)} dr^2 + r^2 ( d\theta^2 + \sin^2\theta d\phi^2 ) ,
\end{equation}
with the metric component function $f(r)$ defined as \cite{HuXR2024a}
\begin{equation}
	f(r) = 1 - \frac{2M}{r} + \frac{Q^2}{r^2} - \frac{aQ^4}{20r^6}
	+ \frac{\lambda_{\text{DM}}}{r} \cdot \ln \frac{r}{|\lambda_{\text{DM}}|} .
	\label{spacetime metric component Euler-Heisenberg}
\end{equation}
In the absence of dark matter (namely $\lambda_{\text{DM}} = 0$), this metric represents an Euler-Heisenberg black hole with nonlinear electrodynamics effects. On the other hand, when $a = 0$, the nonlinear electrodynamics effects vanish and the spacetime gives a charged Reissner-Nordstr\"om (RN) black hole surrounded by PFDM. In the cases of $\lambda_{\text{DM}} = 0$ and $a = 0$, the spacetime recovers to a simple RN black hole spacetime. 

In the presence of PFDM and nonlinear electrodynamics effects, the horizon of charged black hole can be greatly influenced. It is commonly known that the classical charged black hole spacetime (such as RN spacetime) could produce naked singularities when electric charge exceeds its maximal allowed value. The Euler-Heisenberg black hole with a positive nonlinear electrodynamics parameter $a > 0$ and a non-vanishing PFDM parameter $\lambda_{\text{DM}} \ne 0$ would no longer produce naked singularities for arbitrary black hole electric charge (see Appendix \ref{appendix1}). 
However, the Euler-Heisenberg black hole with a negative nonlinear electrodynamics parameter $a < 0$ could still result in naked singularity.
The horizons of Euler-Heisenberg black hole surrounded by PFDM are discussed in Appendix \ref{appendix1}.

As a theoretical investigation, we only consider the dark matter field and nonlinear electromagnetic field in this work. The influences from other complex matter fields (such as disk structure in galaxies composed of luminous stars, dust and plasma in galaxies) are not included in our work. Since the energy-momentum tensor of the dark matter fluid is spherically symmetric, it can form a spherically symmetric halo structure around the central supermassive black hole. In this work, we assume that the event horizon of Euler-Heisenberg black hole (surrounded by PFDM) $r_{H}$, the scale of dark matter halo $r_{\text{halo}}$, and the impact parameter for particle orbits $b$ satisfy
\begin{equation}
	r_{H} \ll b \ll r_{\text{halo}} .
\end{equation}

Furthermore, it should be mentioned that the PFDM model which gives rise to the black hole solution in (\ref{spacetime metric component Euler-Heisenberg}) is a very simplified perfect fluid model. In such simplified model, the energy-momentum tensor of PFDM contains only one free parameter $\lambda_{\text{DM}}$, so the energy density $\rho$ and pressure $P$ of the isotropic dark matter fluid are directly linked with $\lambda_{\text{DM}}$ via expression (\ref{energy-momentum tensor}). The spacetime metric of black holes in this simple dark matter model satisfies $g_{tt} \times g_{rr} = -1$. Recently, some studies have explored more complex perfect fluid models for dark matter. In these dark matter fluid models, the energy-momentum tensor of dark matter fluid contains the energy density $\rho$ and the pressures in radial direction and tangential direction (parameterized by $P_{r}$ and $P_{t}$), and they are connected through the equation of state for dark matter fluids \cite{Cardoso2021,Konoplya2022,Ovgun2025,Daghigh2023ixh}. Especially, some researchers are interested in the Tolman-Oppenheimer-Volkoff (TOV) equation of state for dark matter fluids \cite{Daghigh2023ixh,LiuD2025}. The charged black hole systems solved from these dark matter models are more complex, and the $g_{tt} \times g_{rr} = -1$ relation for spacetime metric is no longer valid. The study of more complex charged black hole solutions from dark matter model is beyond the scope of our present work.   

\section{Theoretical Method} \label{section3}

This section gives descriptions on the theoretical methods used to calculate the gravitational lensing observables, massive object's bound orbit, and black hole shadow. The treatments of particle orbits (which are geodesics) are presented in subsection \ref{section3a}. Schemes to obtain gravitational deflection angle of light and time delay of light for finite distance light source and observer are introduced in subsections \ref{section3b} - \ref{section3c}. The method on the trajectories and precession angle for massive object's bound orbits around black holes is reviewed in subsection \ref{section3d}. Finally, the approach to obtain black hole shadow is given in subsection \ref{section3e}.

\subsection{Geodesics} \label{section3a}

In the gravitational field, massive and massless particles are moving along geodesics in 4 dimensional spacetime. Here we provide a derivation of the reduced geodesic equation, which turns out to be extremely useful in the exploration of gravitational lensing observables. For a spherically symmetric spacetime with the metric form 
\begin{equation}
	d\tau^{2} = f(r)dt^{2} -\frac{1}{f(r)}dr^{2} 
	-r^{2}(d\theta^{2}+\sin^{2}\theta d\phi^{2}) ,
	\label{spacetime metric spherically symmetric}
\end{equation}
the following conserved quantities can be introduced from the symmetry and Killing vectors of spacetime \cite{ChenB2018,Hartle2021}
\begin{subequations}
	\begin{eqnarray}
		J & \equiv & r^{2}\sin^{2}\theta \frac{d\phi}{d\lambda} ,
		\\
		E & \equiv &  f(r)\frac{dt}{d\lambda} ,
		\\
		\epsilon & \equiv & g_{\mu\nu}dx^{\mu}dx^{\nu} 
		= f(r) \bigg( \frac{dt}{d\lambda} \bigg)^2
		- \frac{1}{f(r)} \bigg( \frac{dr}{d\lambda} \bigg)^{2}
		- r^{2} \bigg( \frac{d\theta}{d\lambda} \bigg)^{2} 
		- r^{2}\sin^{2}\theta \bigg( \frac{d\phi}{d\lambda} \bigg)^{2} .
	\end{eqnarray}
\end{subequations}
Here, $\lambda$ is an affine parameter in a particle orbit, $J$ is the conserved angular momentum along the particle orbit, and $E^{2}/2$ can be viewed as the conserved energy along a particle orbit. Specifically, for spherically symmetric spacetime, we can always assume that test particles are moving in the equatorial plane $\theta=\pi/2$, and the reduced differential equation can be obtained using these conserved quantities \cite{ChenB2018,Hartle2021}
\begin{equation}
	\frac{1}{2} \bigg( \frac{dr}{d\lambda} \bigg)^{2} + \frac{1}{2} f(r) \bigg[ \frac{J^{2}}{r^{2}} + \epsilon \bigg]
	= \frac{1}{2} \bigg( \frac{dr}{d\lambda} \bigg)^{2} + V_{\text{eff}}(r)
	= \frac{1}{2}E^{2} . 
	\label{reduced differential equation} 
\end{equation}
Here, $V_{\text{eff}}(r) = \frac{f(r)}{2} \big[ \frac{J^{2}}{r^{2}} + \epsilon \big]$ is the effective potential of test particles moving in a spherically symmetric gravitational field, and $b\equiv |J/E|$ is the impact parameter in a particle orbit (which is also conserved along this orbit). For massless particles traveling along null geodesics, the parameter $\epsilon$ takes the value $\epsilon=0$. For massive particles traveling along timelike geodesics, the parameter $\epsilon$ takes the value $\epsilon=1$. 

\subsection{Gravitational Deflection Angle of Light} \label{section3b}

In astrophysical observations, photons beams emitted from distant luminous sources can be deflected and distorted due to the gravitational field generated by extremely massive celestial bodies (such as supermassive black holes at the center of galaxies, or other huge massive gravitational sources). The gravitational deflection angle, which quantifies the change of tangent direction during the photon propagation, directly reflects the distortion effects induced by a gravitational field on photon (or other particle) trajectories. The larger deflection angle of light corresponds to stronger deflection and distortion effects of gravitation during the photon propagation process.

\begin{figure*}
	\centering
	\includegraphics[width = 0.75 \textwidth]{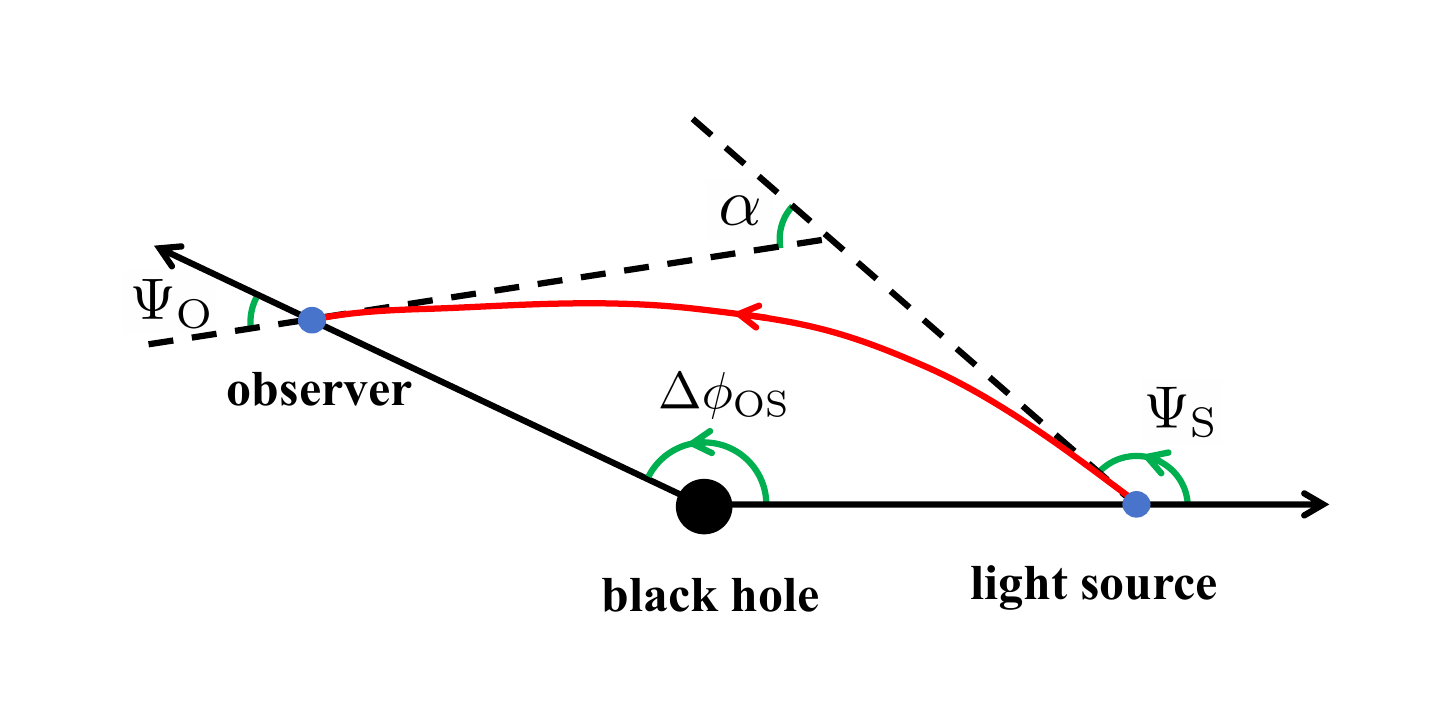}
	\caption{The gravitational deflection angle of light for finite distance light source and observer. This figure illustrate the gravitational deflection angle $\alpha \equiv \Psi_{\text{O}} - \Psi_{\text{S}} - \Delta \phi_{\text{OS}}$ in the thin lens approximation. }
	\label{figure1}
\end{figure*}

In the gravitational lensing observation, when light source and observer are located at a finite distance region to the central black hole (with the radial coordinate $r=r_{\text{S}}$ and $r=r_{\text{O}}$, respectively), the gravitational deflection angle of light is defined through \cite{Ishihara2016}
\begin{equation}
	\alpha \equiv \Psi_{\text{O}} - \Psi_{\text{S}} - \Delta \phi_{\text{OS}} .
	\label{gravitational deflection angle}
\end{equation}
The notations $\Psi_{\text{O}}$, $\Psi_{\text{S}}$ denote the angles between light ray propagation direction and the radial direction (which are measured at the observer's position and source position), and $\Delta \phi_{\text{OS}}$ is the variation of azimuthal angle $\phi$ in the light trajectory. For a spherically symmetric spacetime described by metric (\ref{spacetime metric spherically symmetric}), the angle between light ray propagation direction and the radial direction is uniquely determined by spacetime metric \cite{Ishihara2016}, via
\begin{equation}
	\sin \Psi = \frac{b \cdot \sqrt{f(r)}}{r} .
	\label{angle psi}
\end{equation} 
The definition of gravitational deflection angle of light in expression (\ref{gravitational deflection angle}) for finite distance light source and observer can be understood from an instructive way using thin lens approximation. In the thin lens approximation, the gravitational deflection angle for finite distance light source and observer is illustrated in figure \ref{figure1}. The thin lens approximation assumes that the spacetime is flat everywhere except for the location of gravitational thin lens (the central black hole position). For the quadrangle depicted in figure \ref{figure1}, the exterior angles at each point are $\Psi_{\text{S}}$, $\alpha$, $\pi-\Psi_{\text{O}}$, $\pi - \Delta \phi_{\text{OS}}$, respectively. Under the thin lens approximation, the sum of exterior angles for the quadrangle equals $2\pi$, which justify the definition of gravitational deflection angle $\alpha \equiv \Psi_{\text{O}} - \Psi_{\text{S}} - \Delta \phi_{\text{OS}}$. However, it should be noted that this definition in expression (\ref{gravitational deflection angle}) given by Ishihara \textit{et al.} is not restricted in the thin lens approximation. It is a consistent and proper definition of gravitational deflection angle, which is hold in more general gravitational lensing treatments (it can go beyond the thin lens approximation), see references \cite{Ishihara2016,Ishihara2017,Ono2019,Takizawa2020,Takizawa2020b} for more discussions on this issue.

To calculate the gravitational deflection angle of light, it is necessary to find the variation of azimuthal angle $\phi$ in the photon orbit. In a spherically symmetric spacetime, the variation of azimuthal angle $\phi$ is derived from the reduced differential equation (\ref{reduced differential equation}) for null geodesics 
\begin{equation}
	\frac{d\phi}{dr} = \frac{d\phi}{d\lambda} \cdot \frac{d\lambda}{dr}
	= \pm \frac{1}{ r^{2} \sqrt{\frac{1}{b^{2}}-\frac{f(r)}{r^{2}}} } ,
\end{equation}
where we have used the affine parameter $\epsilon=0$ for massless photon, the conserved energy parameter $E = f(r) \frac{dt}{d\lambda}$, conserved angular momentum $J = r^2 \sin^{2}\theta \frac{d\phi}{d\lambda}$, and the definition of impact parameter $b\equiv |J/E|$. The plus and minus signs $\pm$ can be determined in the following way. When the photon beams travel along a scattering orbit from the source position $r_{\text{S}}$, the radial coordinate $r$ decreases as the azimuthal angle $\phi$ increases, until photon beams reach the closest distance $r=r_0$ to central supermassive black hole. After passing the turning point $r=r_{0}$, the radial coordinate starts to increase with the increasing of azimuthal angle. Thus, we derive the following relation
\begin{subequations}
\begin{eqnarray}
	\frac{d\phi}{dr} = - \frac{1}{ r^{2} \sqrt{\frac{1}{b^{2}}-\frac{f(r)}{r^{2}}} } < 0 , \ \ \ \ \text{from source position $r=r_{\text{S}}$ to tuning point $r=r_{0}$} \ \ \ \ 
	\nonumber
	\\
	\frac{d\phi}{dr} = \frac{1}{ r^{2} \sqrt{\frac{1}{b^{2}}-\frac{f(r)}{r^{2}}} } > 0 . \ \ \ \ \ \ \ \text{from tuning point $r=r_{0}$ to observer position $r=r_{\text{O}}$}  
	\nonumber
\end{eqnarray}
\end{subequations} 
For the closet distance to central black hole in the photon orbit, the derivative $dr/d\lambda$ in equation (\ref{reduced differential equation}) vanishes automatically
\begin{eqnarray}
	\frac{dr}{d\lambda}\bigg|_{r=r_{0}} = 0 
	& \Rightarrow &  
	\frac{1}{2}f(r_{0}) \frac{J^{2}}{r_{0}^{2}} = \frac{1}{2}E^{2} \nonumber
	\\
	& \Rightarrow & b^{2} = \frac{J^{2}}{E^{2}} = \frac{r_{0}^{2}}{f(r_{0})} . \label{impactparameter-closetdistence}
\end{eqnarray}
Given the impact parameter $b$ for a photon orbit, the closet distance $r_{0}$ can be solved from this equation.

With the expression of angle $\Psi$ in equation (\ref{angle psi}) and the variation of azimuthal angle, the gravitational deflection angle of light can be expressed as \cite{Ishihara2016,Hartle2021}
\begin{eqnarray}
	\alpha & = & \Psi_{\text{O}} - \Psi_{\text{S}} - \Delta \phi_{\text{OS}} \nonumber
	\\ 
	& = & \Psi_{\text{O}} - \Psi_{\text{S}} 
	+ \int_{r_{\text{S}}}^{r_{0}} \frac{d\phi}{dr} dr 
	+ \int_{r_{0}}^{r_{\text{O}}} \frac{d\phi}{dr} dr \nonumber
	\\
	& = & \arcsin \bigg( \frac{b \sqrt{f(r_{\text{R}})}}{r_{\text{R}}} \bigg)
	+ \arcsin \bigg( \frac{b \sqrt{f(r_{\text{S}})}}{r_{\text{S}}} \bigg) - \pi 
	+ \int_{r_{0}}^{r_{\text{S}}} \frac{dr}{r^{2}\sqrt{\frac{1}{b^{2}}-\frac{f(r)}{r^{2}}}}
	+ \int_{r_{0}}^{r_{\text{O}}} \frac{dr}{r^{2}\sqrt{\frac{1}{b^{2}}-\frac{f(r)}{r^{2}}}} . 
	\label{gravitational deflection angle for null geodesic} 
\end{eqnarray}
As shown in figure \ref{figure1}, the angle $\Psi$ measured at the position of light source is an obtuse angle such that $\Psi_{\text{S}} = \pi - \arcsin \big( \frac{b \sqrt{f(r_{\text{S}})}}{r_{\text{S}}} \big)$.

\subsection{Time Delay of Light} \label{section3c}

In the gravitational field, the elapse of time (measured by an observer) during the light propagation can be strongly affected by gravitation, and this is the time delay effect of gravitational field. To calculate the time delay of light in gravitational lensing observations, it is necessary to compute the variation of time coordinate $t$ (measured by a distant observer) in the photon orbit. For the massless photon traveling along null geodesics (with $\epsilon=0$), one can derive the time derivative of the radial coordinate from the reduced geodesic equation in (\ref{reduced differential equation}) 
\begin{equation}
	\frac{dr}{dt} = \frac{dr}{d\lambda} \cdot \frac{d\lambda}{dt}
	= \pm f(r) \sqrt{1 - b^{2}\cdot\frac{f(r)}{r^{2}}}  .
\end{equation}
The plus and minus signs $\pm$ can be determined similar to the calculation of gravitational deflection angle. When the photon beams are emitted from the source position $r_{\text{S}}$, the radial coordinate $r$ decreases as time passes, until the photon beams reach the closest distance $r=r_0$ to central supermassive black hole. After passing the turning point $r=r_{0}$, the radial coordinate starts to increase as time advances. So we have the relation
\begin{subequations}
\begin{eqnarray}
	\frac{dt}{dr} = - \frac{1}{ f(r) \sqrt{1 - b^{2}\cdot\frac{f(r)}{r^{2}}} } < 0 , \ \ \ \  \text{from source position $r=r_{\text{S}}$ to tuning point $r=r_{0}$} \ \ \ 
	\nonumber
	\\
	\frac{dt}{dr} = \frac{1}{ f(r) \sqrt{1 - b^{2}\cdot\frac{f(r)}{r^{2}}} } > 0 . \ \ \ \ \ \ \ \text{from tuning point $r=r_{0}$ to observer position $r=r_{\text{O}}$}  
	\nonumber
\end{eqnarray}
\end{subequations} 

In the gravitational lensing observation, when the light source and observer are located at $r=r_{\text{S}}$ and $r=r_{\text{O}}$, the time delay of light during the light propagation process can be expressed as \cite{Hartle2021,Weinberg1972}
\begin{eqnarray}
	\Delta T = T - T_{0} & = & \int_{r_{\text{S}}}^{r_{0}} \frac{dt}{dr} dr + \int_{r_{0}}^{r_{\text{O}}} \frac{dt}{dr} dr 
	- T_{0} \nonumber
	\\
	& = & -\int_{r_{\text{S}}}^{r_{0}} \frac{dr}{f(r)\sqrt{1-\frac{b^{2} f(r)}{r^{2}}}}
	+ \int_{r_{0}}^{r_{\text{O}}} \frac{dr}{f(r)\sqrt{1-\frac{b^{2} f(r)}{r^{2}}}}
	- T_{0} \nonumber
	\\
	& = & \int_{r_{0}}^{r_{\text{S}}} \frac{dr}{f(r)\sqrt{1-\frac{b^{2} f(r)}{r^{2}}}}
	+ \int_{r_{0}}^{r_{\text{O}}} \frac{dr}{f(r)\sqrt{1-\frac{b^{2} f(r)}{r^{2}}}}
	- \sqrt{r_{\text{S}}^{2}-r_{0}^{2}} - \sqrt{r_{\text{O}}^{2}-r_{0}^{2}} ,
	\label{time delay for null geodesic}
\end{eqnarray}
with $T_{0} =\sqrt{r_{\text{S}}^{2}-r_{0}^{2}}+\sqrt{r_{\text{O}}^{2}-r_{0}^{2}}$ to be the time period during the light propagation process without the gravitational field. In the expression (\ref{time delay for null geodesic}), it is obvious that the time delay $\Delta T$ increases monotonically as the coordinates of light source and observer $r_{\text{S}}$, $r_{\text{O}}$ increase. Additionally, in the integration, the turning point $r=r_{0}$ is solved through equation (\ref{impactparameter-closetdistence}). 

Furthermore, it should be addressed that the time delay of light calculated using expression (\ref{time delay for null geodesic}) is the time delay measured by the conventional static time coordinate $t$ in spacetime metric expression (\ref{spacetime metric spherically symmetric}). For an asymptotically flat spacetime, the static time coordinate corresponds to the time measured by a very distant observer (at the asymptotically flat region). On the other hand, if one wants to calculate the time delay measured by the proper time of a finite distance static observer, a redshift factor should be taken into account
\begin{equation}
	\Delta T_{\text{static observer's proper time unit}} = \frac{\Delta T}{\text{redshift factor}}
	= \frac{\Delta T}{\sqrt{f(r_{\text{O}})}} ,
\end{equation}	 
with $r_{\text{O}}$ to be the position of the finite distance static observer.

\subsection{Massive Object's Orbits and Precession Angle} \label{section3d}

Besides the gravitational deflection angle and time delay of light discussed in subsections \ref{section3b}-\ref{section3c} (which is relevant to a scattering orbit for massless photons), the massive object's orbits are also greatly influenced by the gravitational field, both for bound orbits and scattering orbits. Notably, massive object's bound orbits may exhibit precession. In astrophysical observations, the precession of bound orbits is frequently used to test the gravitational field around the supermassive black hole in a galaxy center (such as the observations relevant to Sgr*A in Milky Way Galaxy).

For a spherically symmetric black hole spacetime, we can always restrict the motion of massive object's in the equatorial plane. To calculate the massive object's orbit in the equatorial plane, it is necessary to resort to the reduced geodesic equation (\ref{reduced differential equation}) and the definition of conserved momentum
\begin{subequations}
\begin{eqnarray}
	\bigg( \frac{dr}{d\tau} \bigg)^{2} 
	& = & E^{2} - V_{\text{eff}}(r) 
	  =   E^{2} - f(r) \bigg[ \frac{J^{2}}{r^{2}} + 1 \bigg] ,
	\label{equation_dr_dtau}
	\\
	\frac{d\phi}{d\tau} & = & \frac{J}{r^{2}} , \label{equation_dphi_dtau}
\end{eqnarray}
\end{subequations}
where we have used the proper time as the affine parameter for massive objects traveling along timelike geodesics. The parameter $\epsilon$ takes the value $\epsilon = 1$ for timelike geodesics. Substitute the expression (\ref{equation_dphi_dtau}) into equation (\ref{equation_dr_dtau}) and cancel the affine parameter $\tau$, the geodesic equation can be further reduced to
\begin{eqnarray}
	&& 
	\bigg( \frac{dr}{d\phi} \frac{J}{r^{2}} \bigg)^{2} 
	= E^{2} - V_{\text{eff}}(r) 
	= E^{2} - f(r) \bigg[ \frac{J^{2}}{r^{2}} + 1 \bigg]
	\nonumber 
	\\
	& \Rightarrow &
	\bigg( \frac{du}{d\phi} \bigg)^{2} 
	= \frac{1}{b^{2}} - F(u) \bigg[ u^{2} + \frac{1}{J^{2}} \bigg] ,
\end{eqnarray}
where we have used a new variable $u=1/r$ in the geodesic equation. Differentiating each side of this equation with respect to $\phi$, a second-order differential equation for massive object's orbit can be obtained \cite{ChenB2018,Hartle2021}
\begin{equation}
	\frac{d^{2}u}{d\phi^{2}} 
	= - \frac{1}{2} \cdot \frac{dF(u)}{du} \cdot \bigg[ u^{2}+\frac{1}{J^{2}} \bigg]
	  - F(u) \cdot u \ .
	\label{orbit differential equation}
\end{equation}
The function $F(u) = f(1/r)$ is the spacetime metric component expressed using the new variable $u=1/r$. Eventually, the massive object's orbit is determined by numerically solving the differential equation (\ref{orbit differential equation}). 

\begin{figure*}
	\centering
	\includegraphics[width=0.7\textwidth]{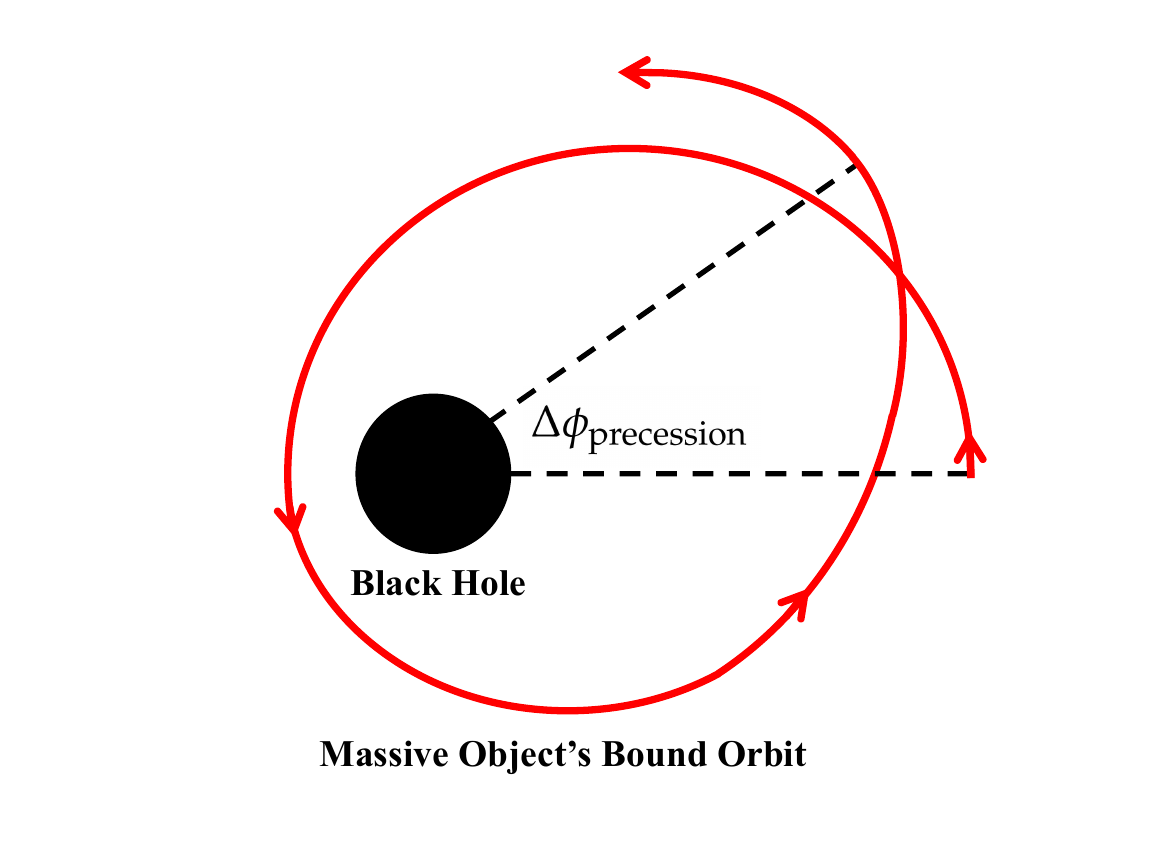}
	\caption{A schematic plot of the precession of azimuthal angle for a massive object's bound orbit around the central black hole. }
	\label{figure1b}
\end{figure*}

An important feature of the massive object's bound object is the precession of azimuthal angle $\phi$, which is illustrated in figure \ref{figure1b}. Given a bound timelike geodesic orbit for a massive object, there are maximal and minimal distance positions to central black hole in this orbit, denoting as $r_{\text{max}}$ and $r_{\text{min}}$. The precession of azimuthal angle $\phi$ in an orbit period can be calculated as \cite{Hartle2021,Weinberg1972}
\begin{eqnarray}
	\Delta \phi_{\text{precession}} 
	& = & \int_{r_{\text{min}}}^{r_{max}} \frac{d\phi}{dr} dr 
	+ \int_{r_{\text{max}}}^{r_{min}} \frac{d\phi}{dr} dr 
	- 2\pi \nonumber
	\\
	& = & 2\int_{r_{\text{min}}}^{r_{max}} 
	\frac{dr}{ r^{2} \sqrt{\frac{1}{b^{2}}-\frac{f(r)}{r^{2}}-\frac{f(r)}{J^{2}}}}  
	- 2\pi .
	\label{precession angle}
\end{eqnarray}
In the derivation of precession angle $\Delta \phi_{\text{precession}}$, we have used the derivative of azimuthal angle with respect to radial coordinate for a timelike geodesic orbit  
\begin{equation}
	\frac{d\phi}{dr} = \frac{d\phi}{d\tau} \cdot \frac{d\tau}{dr} 
	= \pm \frac{1}{ r^{2} \sqrt{\frac{1}{b^{2}}-\frac{f(r)}{r^{2}}-\frac{f(r)}{J^{2}}} } .
\end{equation}
The plus and minus signs $\pm$ can be determined as follows. When a massive object begins to move along timelike geodesics from the maximal distance position $r_{\text{max}}$, the radial coordinate $r$ decreases as the azimuthal angle $\phi$ increases, until the massive object reaches the minimal distance position $r=r_{\text{min}}$ to central supermassive black hole. After passing the minimal distance position $r=r_{\text{min}}$, the massive object's radial coordinate starts to increase as the azimuthal angle increases. Thus, we can get the relation
\begin{subequations}
\begin{eqnarray}
	\frac{d\phi}{dr} = - \frac{1}{ r^{2} \sqrt{\frac{1}{b^{2}}-\frac{f(r)}{r^{2}}-\frac{f(r)}{J^{2}}} } < 0 , \ \ \ \text{from maximal distance $r=r_{\text{max}}$ to minimal distance $r=r_{\text{min}}$} 
	\nonumber
	\\
	\frac{d\phi}{dr} = \frac{1}{ r^{2} \sqrt{\frac{1}{b^{2}}-\frac{f(r)}{r^{2}}-\frac{f(r)}{J^{2}}} } > 0 . \ \ \ \ \ \text{from minimal distance $r=r_{\text{min}}$ to maximal distance $r=r_{\text{max}}$}  
	\nonumber
\end{eqnarray}
\end{subequations} 
Furthermore, the maximal and minimal distance positions to the central black hole in a massive object's bound orbit can be determined via 
\begin{equation}
	\frac{dr}{d\phi} \bigg|_{r=r_{\text{max}},\ r=r_{\text{min}}} 
	=  r^{2} \sqrt{\frac{1}{b^{2}}-\frac{f(r)}{r^{2}}-\frac{f(r)}{J^{2}}}  \bigg|_{r=r_{\text{max}},\ r=r_{\text{min}}} = 0 .
\end{equation}

\subsection{Black Hole Shadow} \label{section3e}

This subsection presents the theoretical method to obtain the black hole shadow size. The critical shadow radius of a black hole is determined by the unstable photon sphere, which can be calculated from the effective potential of photons moving in a gravitational field, or through a geometric approach proposed in a recent works \cite{QiaoCK2022a}. In this work, we calculate the position of unstable photon sphere and the black hole shadow radius based on the geometric approach, in which the photon sphere and its stability are uniquely determined from Gaussian curvature and geodesic curvature in the optical geometry of black hole spacetime. Firstly, any photon orbits maintain their geodesic nature when they are transformed into optical geometry, so the geodesic curvature for a photon sphere vanishes. This geodesic curvature condition for photon sphere is completely equivalent to find the local extreme point of effective potential \cite{QiaoCK2022a,QiaoCK2022b,QiaoCK2024b}
\begin{equation} 
	\kappa_{g}(r=r_{\text{ph}})  =  0  
	\ \ \Leftrightarrow \ \
	\frac{dV_{\text{eff}}(r)}{dr} \bigg|_{r=r_{\text{ph}}} = 0 .
	\label{photon sphere}
\end{equation}
Furthermore, the unstable photon spheres (or stable photon spheres) are determined by the negative (or positive) sign of Gaussian curvature, which is equivalent to find the local maximum (or local minimum) points of effective potential \cite{QiaoCK2022a,QiaoCK2022b,QiaoCK2024b}
\begin{subequations} 
	\begin{eqnarray}
		\mathcal{K}(r=r_{\text{unstable}}) < 0 
		\ \ & \Leftrightarrow & \ \ 
		\frac{d^{2}V_{\text{eff}}(r)}{dr^{2}} \bigg|_{r=r_{\text{unstable}}} < 0 ,
		\label{unstable photon sphere}
		\\
		\mathcal{K}(r=r_{\text{stable}}) > 0
		\ \ & \Leftrightarrow & \ \ 
		\frac{d^{2}V_{\text{eff}}(r)}{dr^{2}} \bigg|_{r=r_{\text{stable}}} > 0 .
	\end{eqnarray}
\end{subequations}
The absolute apparent size of black hole shadow viewed by an observer located at 
infinity (the asymptotically flat region) is characterized by the critical shadow radius. With the unstable photon sphere obtained in equations (\ref{photon sphere}) and (\ref{unstable photon sphere}), the critical shadow radius of a black hole can be calculated through the relation
\begin{equation} 
	r_{sh} = b_{\text{critical}} 
	= \frac{r_{\text{unstable}}}{\sqrt{f(r=r_{\text{unstable}})}} .
	\label{critical shadow radius}
\end{equation}

\section{Results and Discussions} \label{section4}

This section presents results on the gravitational lensing observables for Euler-Heisenberg black hole surrounded by PFDM, including the gravitational deflection angle of light and time delay of light. The trajectory and precession angle of massive object's bound orbits are also presented in this section. These quantities are calculated numerically using expressions (\ref{gravitational deflection angle for null geodesic}), (\ref{time delay for null geodesic}) and (\ref{orbit differential equation}). Furthermore, inspired by the rapid development of black hole shadow research since the capture of black hole images by ETH, we also present a discussion on the black hole shadow for Euler-Heisenberg black hole surrounded by PFDM. 

\subsection{Gravitational Deflection Angle of Light} \label{section4a}

\begin{figure}[t]
	\centering
	\includegraphics[width = 0.495 \textwidth]{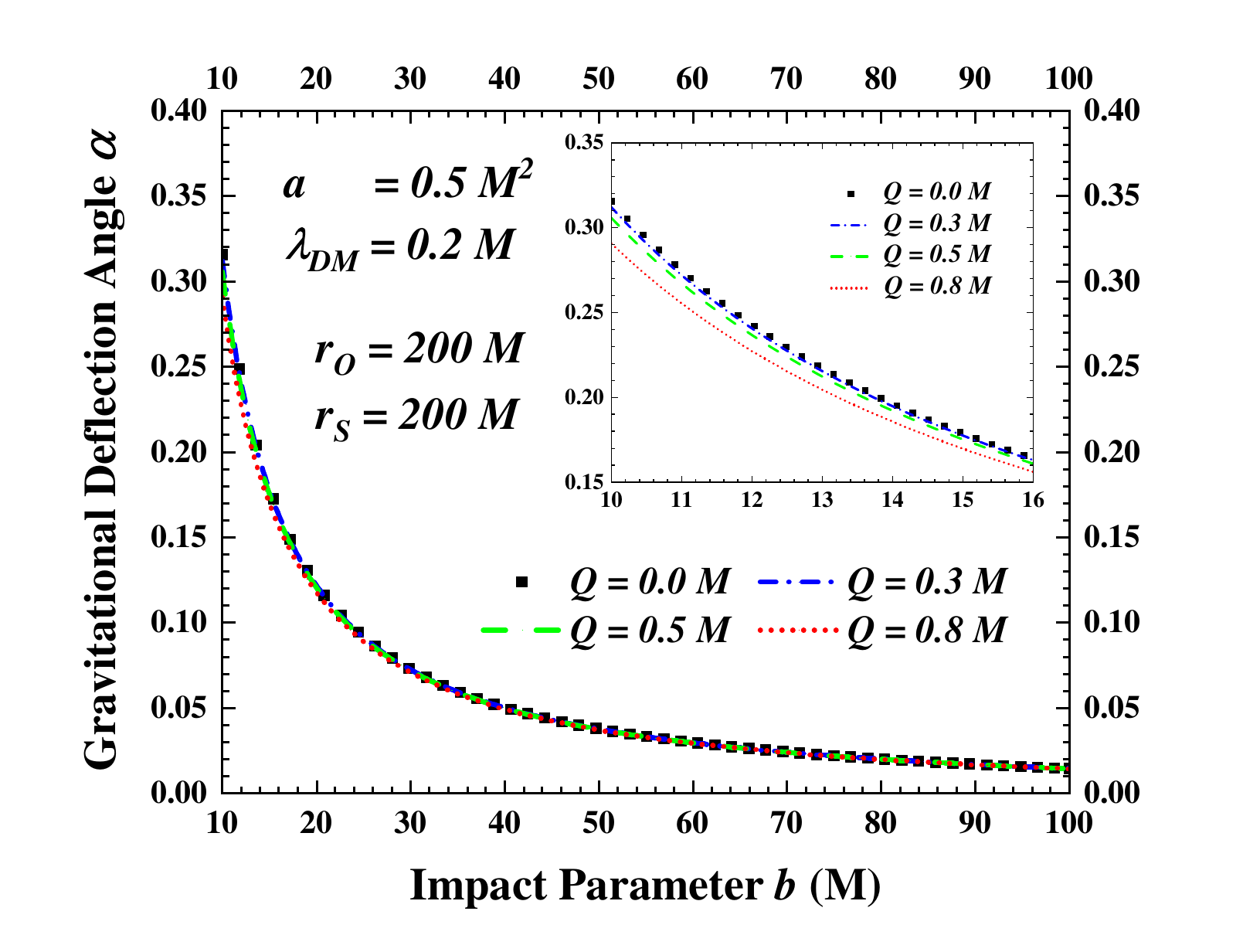}
	\includegraphics[width = 0.495 \textwidth]{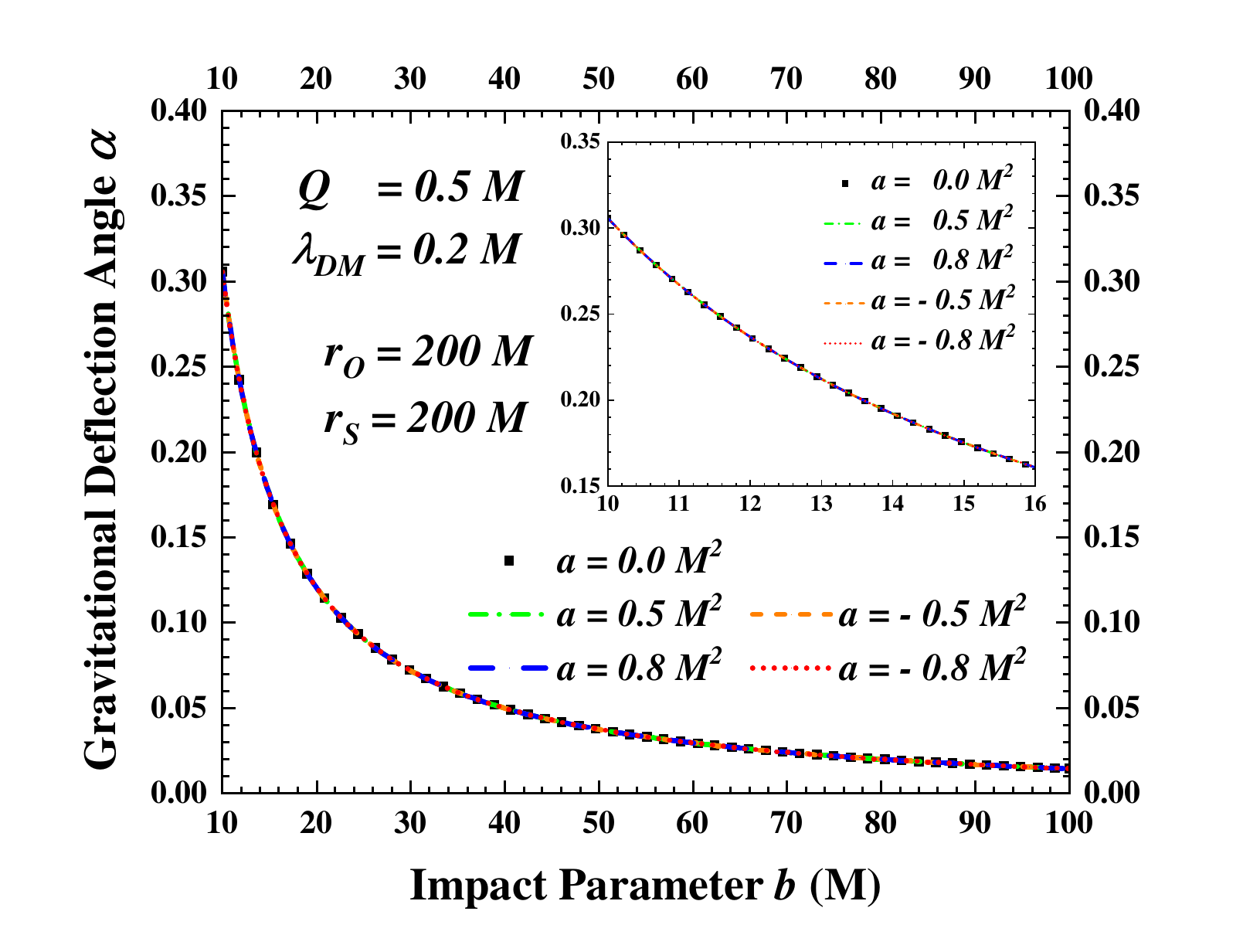}
	\includegraphics[width = 0.495 \textwidth]{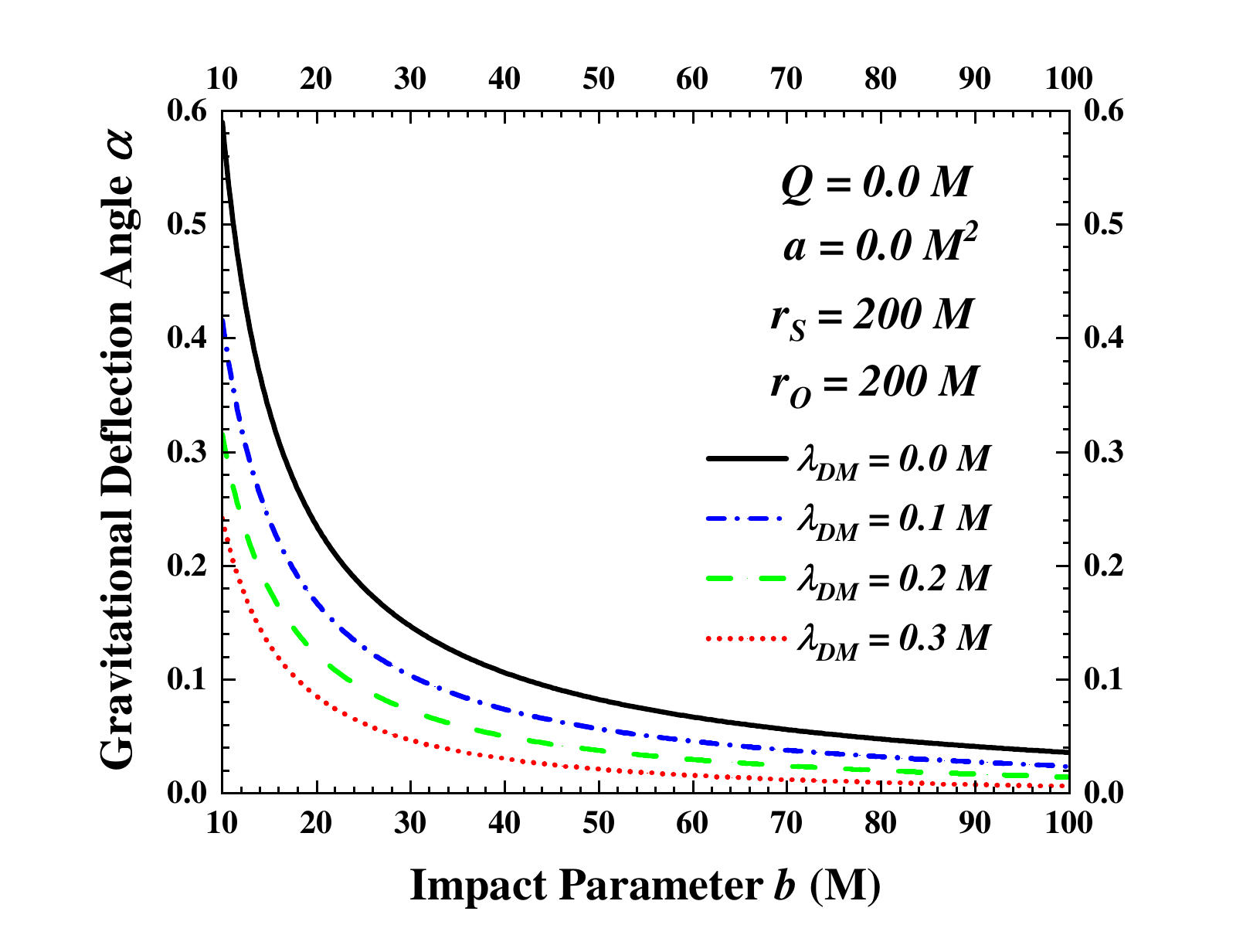}
	\includegraphics[width = 0.495 \textwidth]{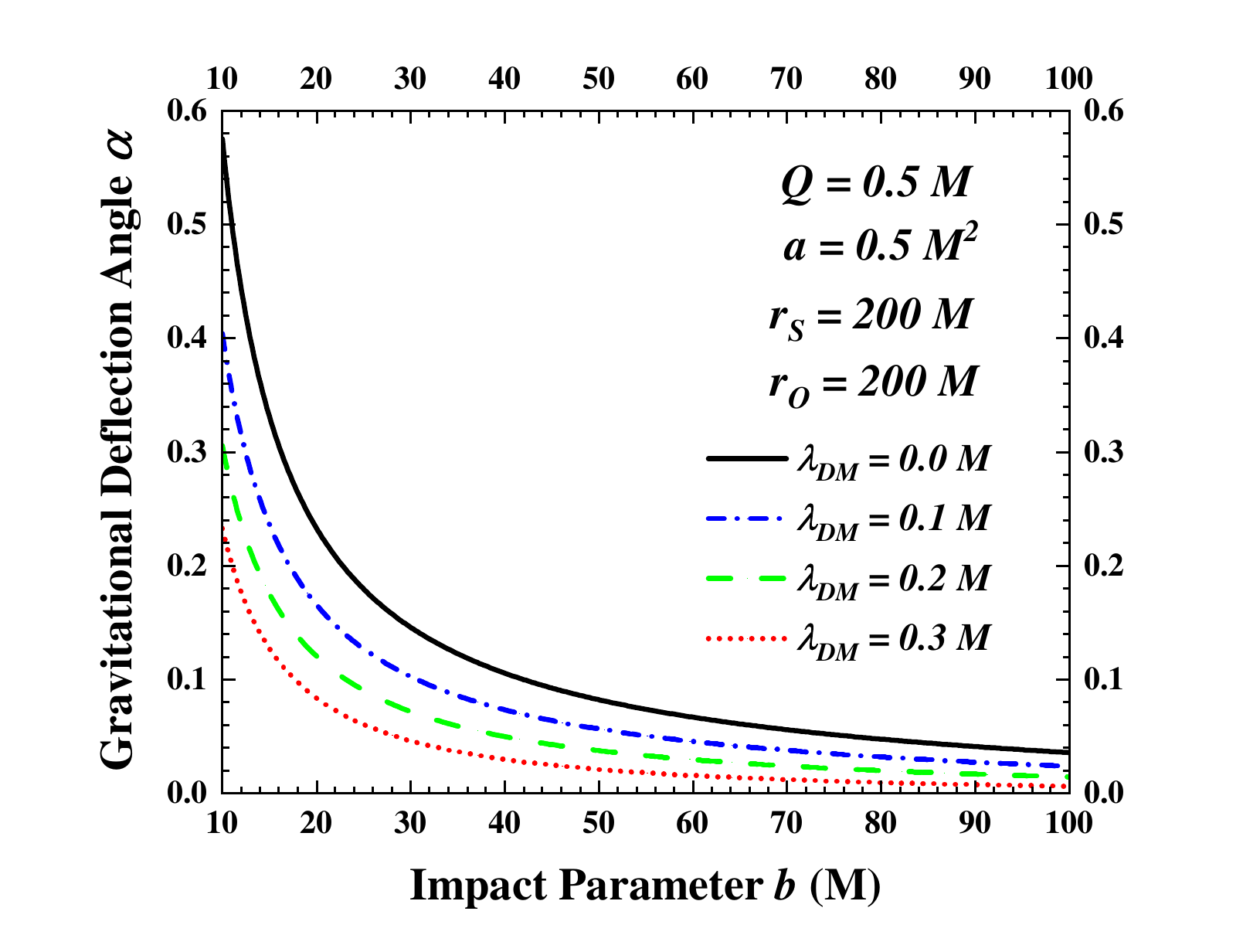}
	\caption{The gravitational deflection angle of light for Euler-Heisenberg black hole surrounded by PFDM. (\textbf{a}) The upper left panel exhibits the influences from black hole electric charge $Q$. (\textbf{b}) The upper right panel shows the influences from nonlinear electrodynamics effects / QED effects. (\textbf{c}) The lower left panel highlights the influences from dark matter in the absence of electric charge and nonlinear electrodynamics effects. (\textbf{d}) The lower right panel highlights the influences from dark matter in the presence of nonlinear electrodynamics effects. In all panels, the horizontal axis labels the impact parameter in photon orbit. The positions of light source and observer are located at a certain distance $r_{\text{O}} = r_{\text{S}} = 200 M$. \label{figure gravitational deflection angle}}
\end{figure}   

The gravitational deflection angles of light for Euler-Heisenberg black hole surrounded by PFDM are given in figure \ref{figure gravitational deflection angle}. They are calculated numerically from expression (\ref{gravitational deflection angle for null geodesic}) for light source and observer located at a finite distance region. The multiple panels of figure \ref{figure gravitational deflection angle} highlight the influences of black hole electric charge, nonlinear electrodynamics effects and dark matter effects on gravitational deflection angles. The upper left and upper right panels highlight the influences from black hole electric charge $Q$ and the nonlinear electrodynamics effects / QED effects (parameterized by $a$), while the lower panels highlight the influences from dark matter parameterized by $\lambda_{\text{DM}}$. To show the gravitational deflection in strong gravitational field cases, the minimal value of impact parameter is chosen as $b \sim 10 M$, which indicates the photon orbit is very close to the black hole critical shadow radius.

The upper left panel of figure \ref{figure gravitational deflection angle} suggests that, for a constant PFDM parameter and nonlinear electrodynamics parameter, the gravitational deflection angles are slightly reduced when black hole's electric charge becomes larger. But the effect coming from electric charge is small such that we must add a magnified subplot into upper left panel to separate the various curves corresponding to different black hole charges. The upper right panel of figure \ref{figure gravitational deflection angle} suggests that the nonlinear electrodynamics parameter has no obvious influences on the gravitational deflection angles, even if a magnified subplot is employed. Comparing the upper and lower panels of figure \ref{figure gravitational deflection angle}, we can observe that both electric charge and nonlinear electrodynamics have minor impacts on gravitational deflection angle, while the most significant influences coming from the PFDM. 
From the lower panels of figure \ref{figure gravitational deflection angle}, it can be clearly manifested that the variation of PFDM parameter could dramatically change the gravitational deflection angle of light for charged Euler-Heisenberg black hole. 
Notably, the gravitational deflection angle reduces rapidly as the increasing of PFDM parameter $\lambda_{\text{DM}}$. This can be explained from the spacetime metric in expression (\ref{spacetime metric component Euler-Heisenberg}) for Euler-Heisenberg black hole surrounded by PFDM. The dark matter effects contribute to the spacetime metric function $f(r)$ via the $-\lambda_{\text{DM}}/r$ term (if we naively omit the slowly varying logarithmic term $\ln\frac{r}{|\lambda_{\text{DM}}|}$ for simplicity), which is similar to the black hole's mass term $2M/r$. This indicates that the PFDM parameter could play a role of ``effective mass'' in the spacetime metric \footnote{Strictly specking, because of the presence of logarithmic term, this ``effective mass'' should be $m_{\text{eff}}(r) = \lambda_{\text{DM}}\ln\frac{r}{|\lambda_{\text{DM}}|}$, rather than $m_{\text{eff}} = \lambda_{\text{DM}}$.}. A positive PFDM parameter ($\lambda_{\text{DM}}>0$) contributes to a negative effective mass in the spacetime metric, which explains the rapid drop of the gravitational deflection angle when PFDM parameter becomes larger. Furthermore, combining the four panels of figure \ref{figure gravitational deflection angle}, it is also interesting to find that the gravitational deflection angles of light for Euler-Heisenberg black hole surrounded by PFDM (with nonzero electric charge $Q \neq 0$ and dark matter parameter $\lambda_{\text{DM}} \neq 0$) are smaller than those in Schwarzschild black hole cases (which correspond to the black solid line in the lower left panel of figure \ref{figure gravitational deflection angle}).  

\subsection{Time Delay of Light} \label{section4b}

\begin{figure}[t]
	\centering
	\includegraphics[width = 0.495 \textwidth]{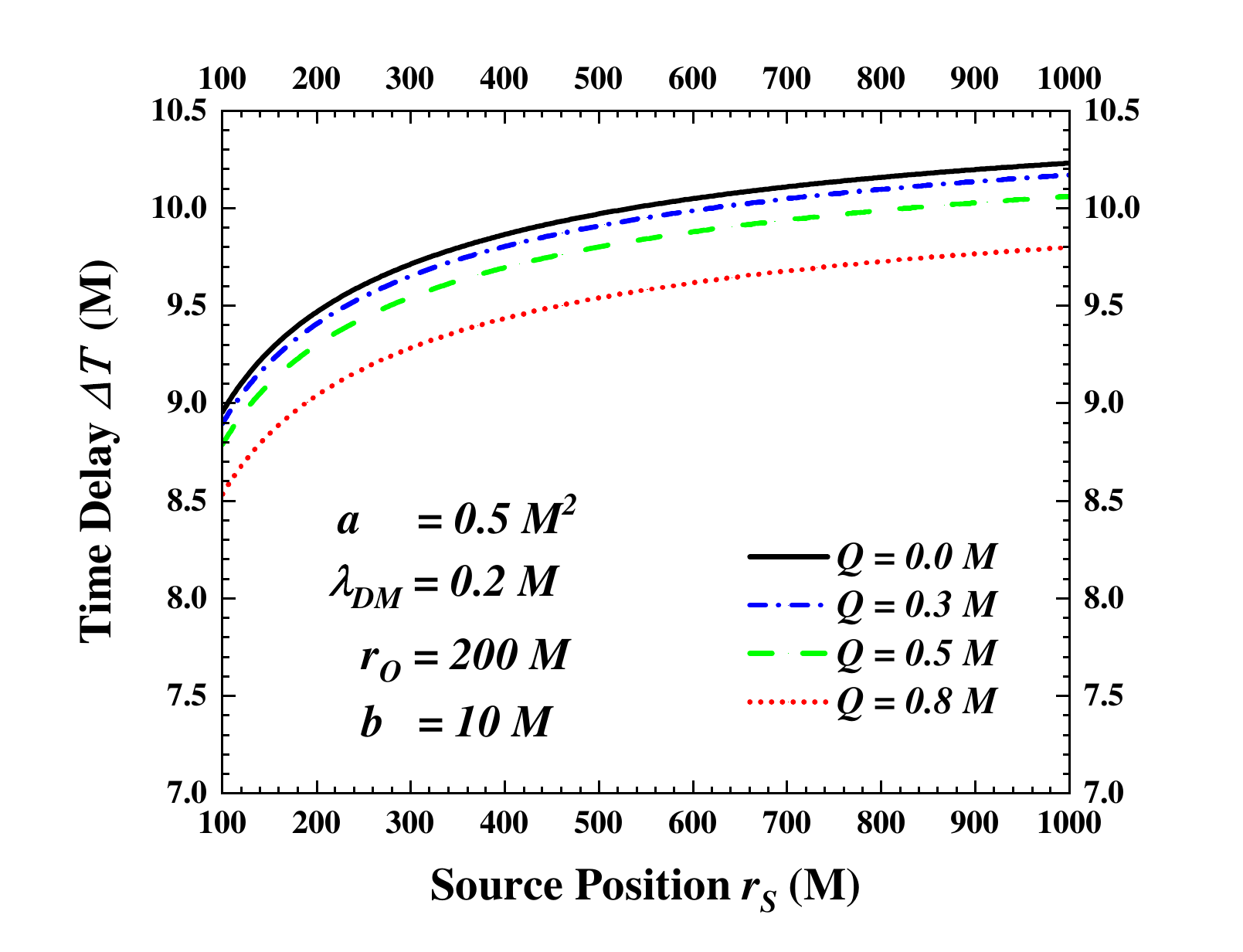}
	\includegraphics[width = 0.495 \textwidth]{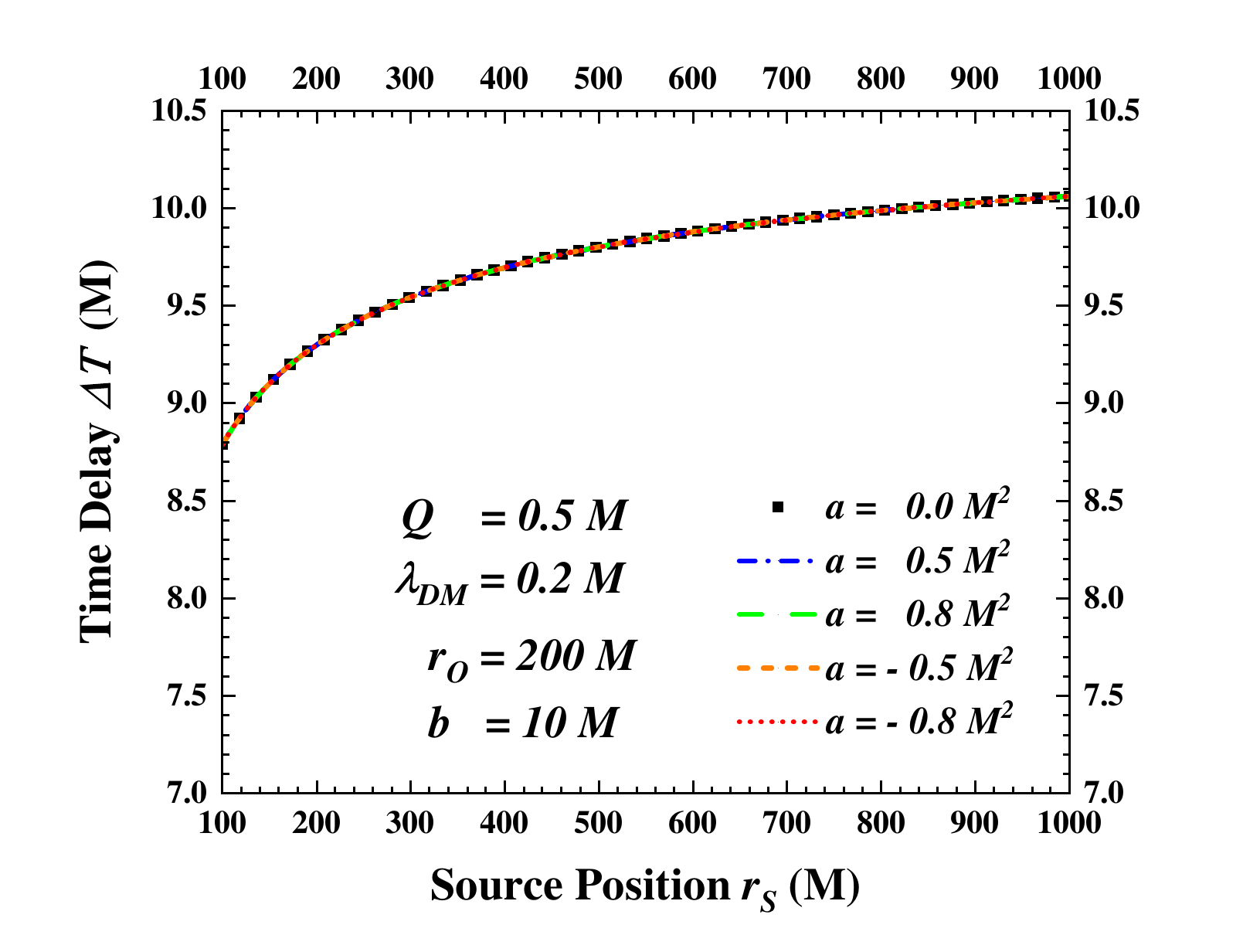}
	\includegraphics[width = 0.495 \textwidth]{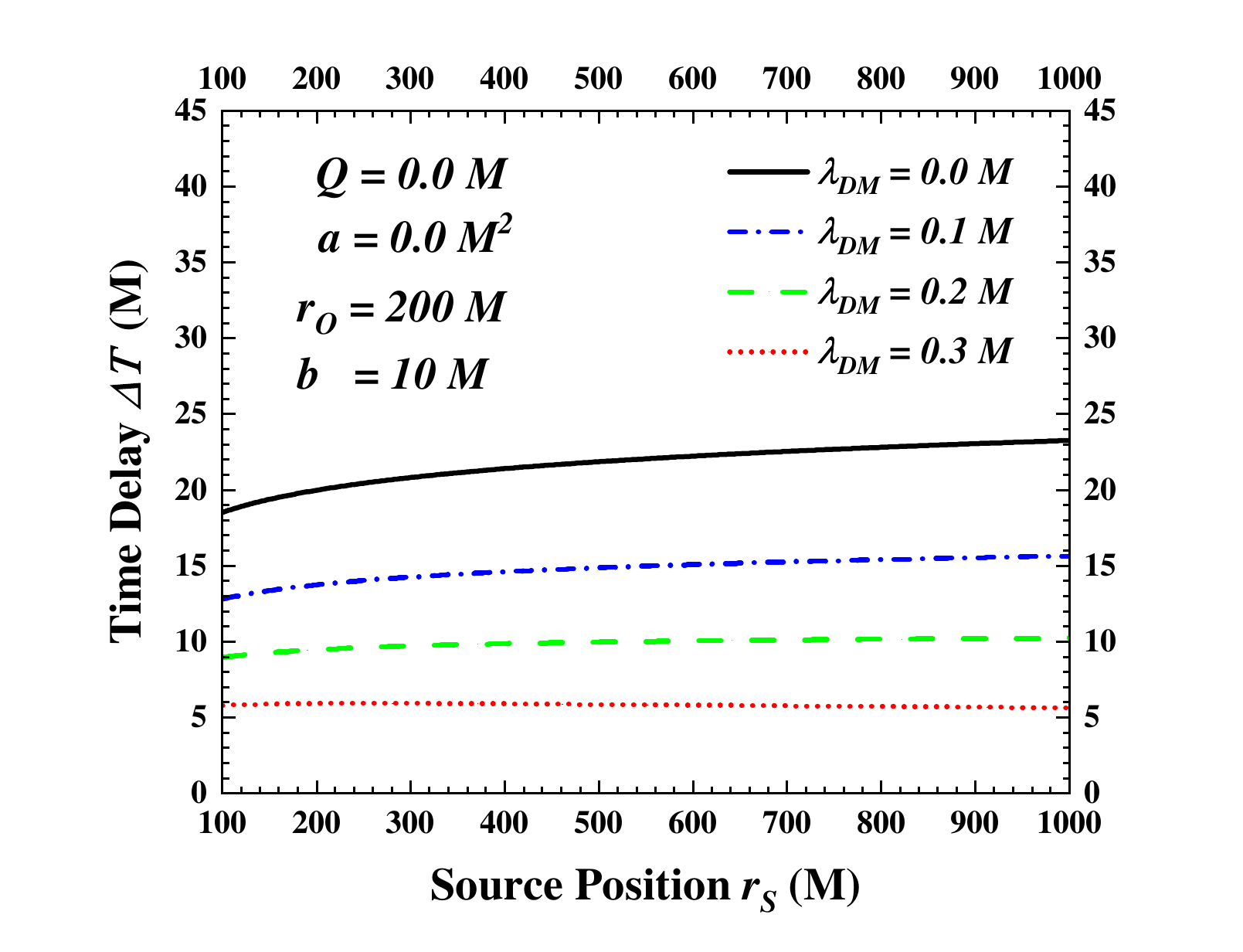}
	\includegraphics[width = 0.495 \textwidth]{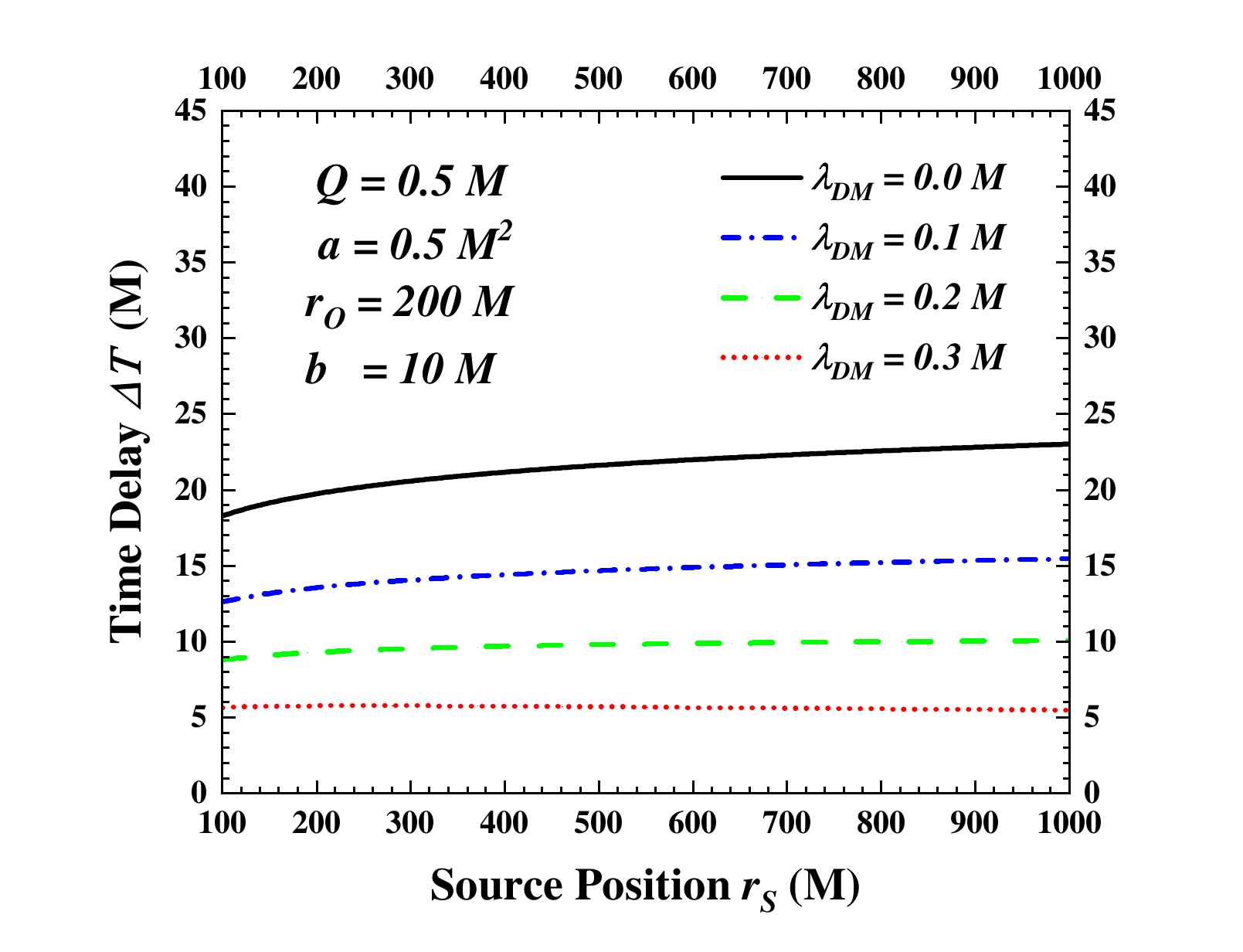}
	\caption{The gravitational time delay of light for Euler-Heisenberg black hole surrounded by PFDM. (\textbf{a}) The upper left panel shows the influences from black hole electric charge $Q$. (\textbf{b}) The upper right panel exhibits the influences from nonlinear electrodynamics effects / QED effects. (\textbf{c}) The lower left panel highlights the influences from dark matter in the absence of black hole charge and nonlinear electrodynamics effects. (\textbf{d}) The lower right panel highlights the influences from dark matter in the presence of nonlinear electrodynamics effects. In all panels, the horizontal axis labels the position of light source, and the vertical axis labels the time delay of light $\Delta T$ measured in unit of black hole mass. 
	\label{figure time delay}}
\end{figure}

The time delay results for Euler-Heisenberg black hole surrounded by PFDM are presented in figure \ref{figure time delay}. This figure plots the time delay of light measured by the conventional static time coordinate $t$, which is calculated numerically from expression (\ref{time delay for null geodesic}) for light source and observer located at a finite distance region. In this figure, the time delays of light varied with respect to light source position $r_{\text{S}}$ are plotted. A similar plot can be carried out for time delays varied with respect to observer position $r_{\text{O}}$, while these results exhibit very similar behaviors with those in figure \ref{figure time delay}. So the plot of time delays varied with respect to observer position is not given in this work. The multiple panels of figure \ref{figure time delay} highlight the influences coming from various black hole parameters. The upper left and upper right panels highlight the influences from black hole electric charge $Q$ and the nonlinear electrodynamics effects (or QED effects) parameterized by $a$, while the lower panels highlight the influences from dark matter parameterized by $\lambda_{\text{DM}}$. In all panels, the time delays of light all grow with the increasing of light source radius $r_{\text{S}}$, confirming that the time delay of light $\Delta T$ calculated using expression (\ref{time delay for null geodesic}) is indeed a monotonically increasing function of light source radius $r_{\text{S}}$ (or observer radius $r_{\text{O}}$).

Comparing the time delay results in figure \ref{figure time delay} and the gravitational deflection results in figure \ref{figure gravitational deflection angle}, it is easy to observe that the black hole parameters (including the black hole electric charge $Q$, nonlinear electrodynamics parameter $a$, PFDM parameter $\lambda_{\text{DM}}$) influence time delays of light in a similar trend as those for gravitational deflection angles. Firstly, Euler-Heisenberg black hole with a larger electric charge results in a reduction of time delay, as shown in the upper left panel of figure \ref{figure time delay}. Secondly, the nonlinear electrodynamics effects (parameterized by $a$) do not have obvious influences on time delay results, as indicated in the upper right panel of figure \ref{figure time delay}. Thirdly, the most significant impacts on time delay come from the PFDM, which are much greater than those from electric charge and nonlinear electrodynamics. From the lower panels of figure \ref{figure time delay}, the time delay of light changes rapidly as the PFDM parameter varies. Particularly, the time delay of light could undergo a rapid reduction for larger PFDM parameters, compared with those in the absence of dark matter (with $\lambda_{\text{DM}}=0$). This can be explained by the ``effective mass'' role of PFDM played in the spacetime metric, the same as we have elaborated in subsection \ref{section4a} for gravitational deflection angle cases. A positive PFDM parameter $\lambda_{\text{DM}}>0$ contributes to a negative effective mass in the spacetime metric, which explains the rapid reduction of time delay for larger PFDM parameter values. Furthermore, combining the four panels of figure \ref{figure time delay}, it is worth noting that the time delays of light for Euler-Heisenberg black hole surrounded by PFDM (with nonzero electric charge $Q \neq 0$ and dark matter parameter $\lambda_{\text{DM}} \neq 0$) are smaller than those in Schwarzschild black hole cases (which correspond to the black solid line in the lower left panel of figure \ref{figure time delay}).

\subsection{Massive Object's Bound Orbits and Precession Angle} \label{section4c} 

\begin{figure}
	\centering
	\includegraphics[width = 7.5 cm]{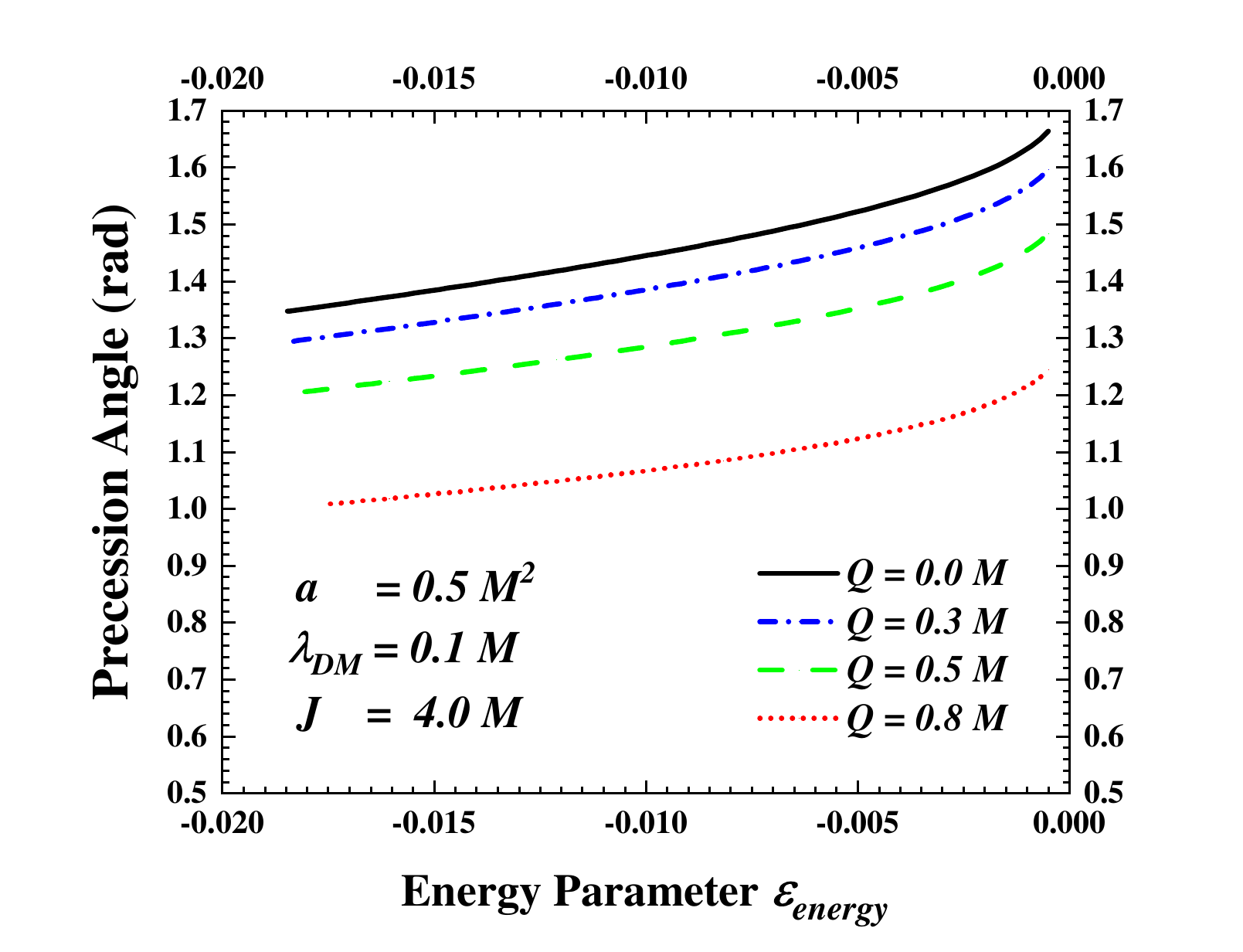}
	\includegraphics[width = 7.5 cm]{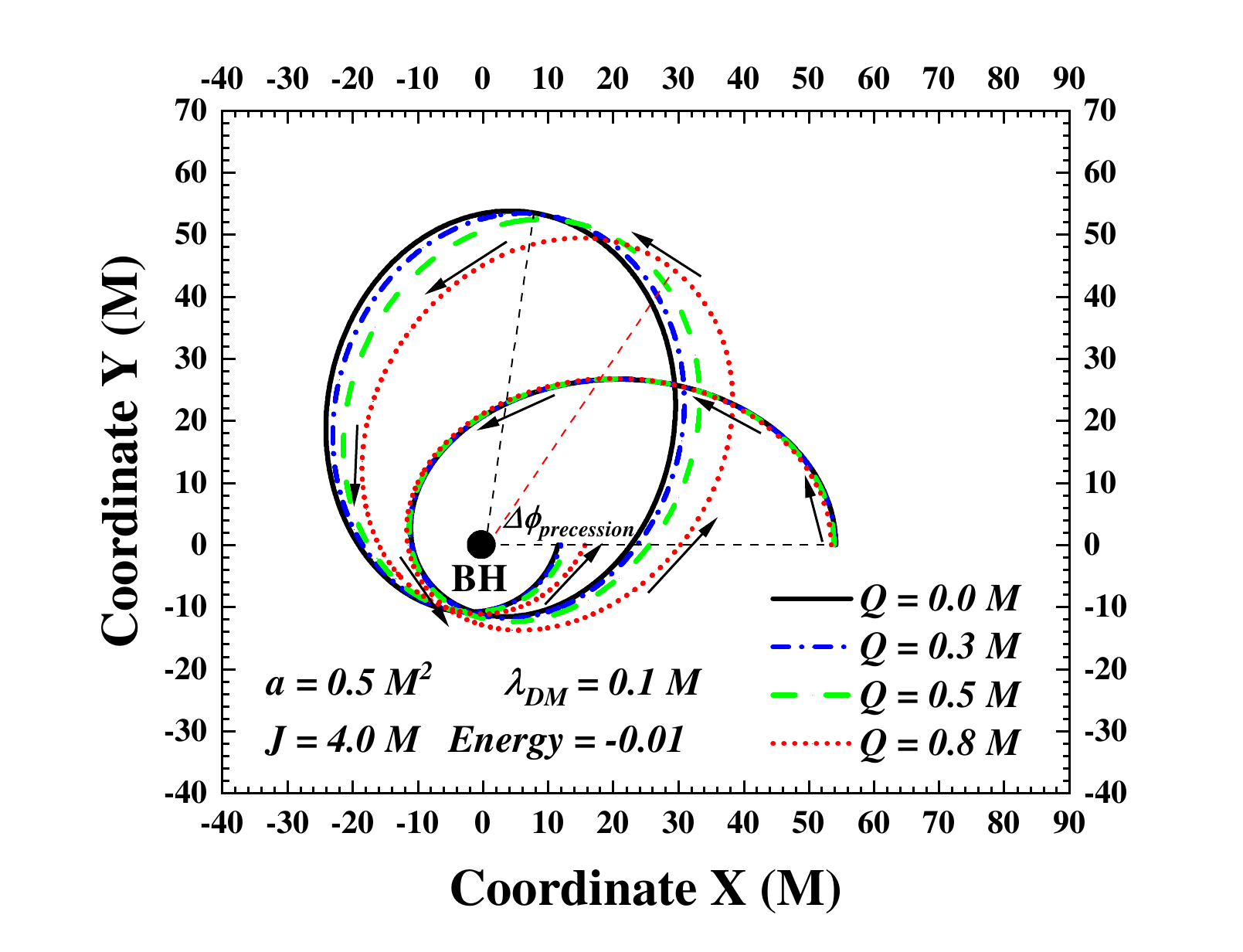}
	\includegraphics[width = 7.5 cm]{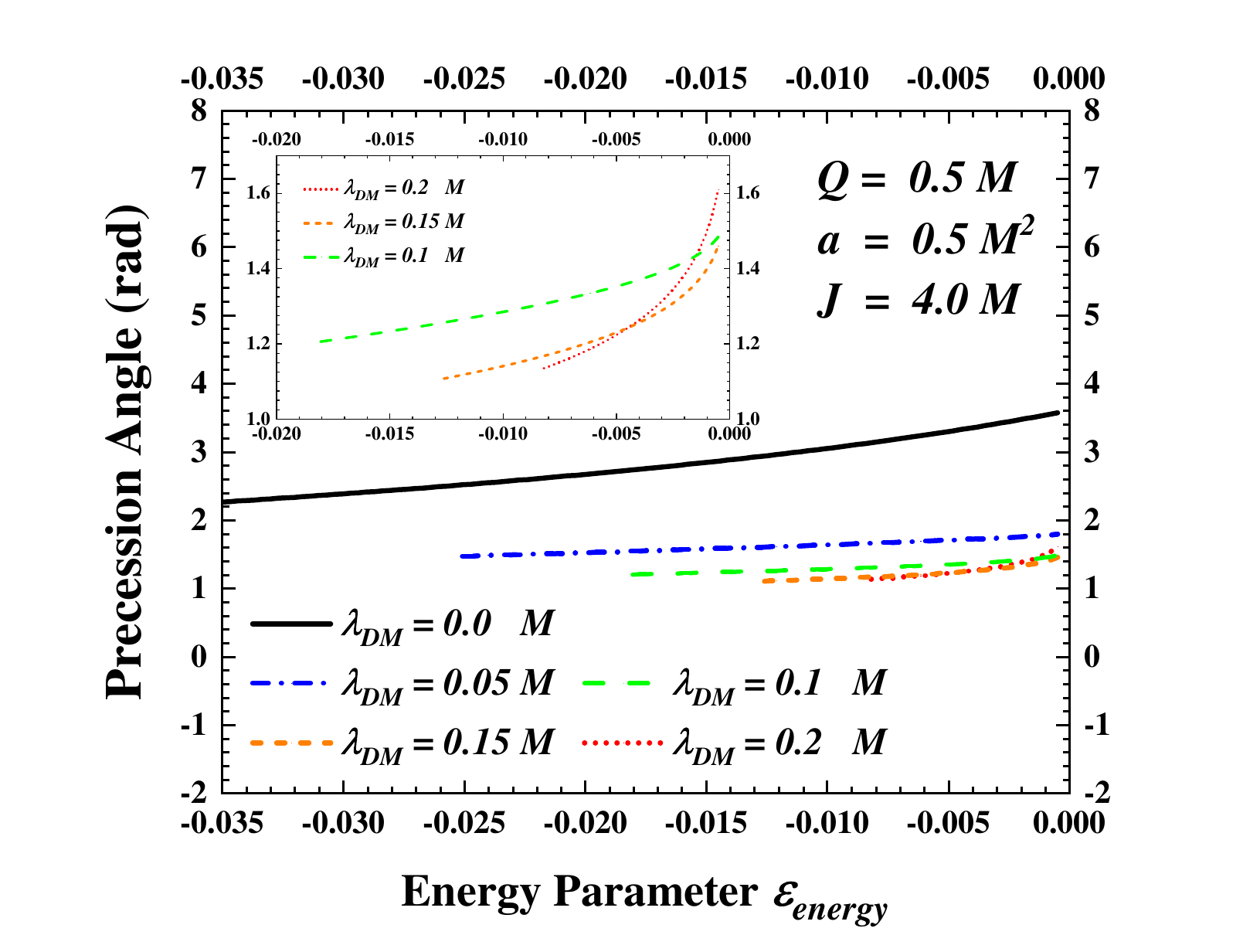}
	\includegraphics[width = 7.5 cm]{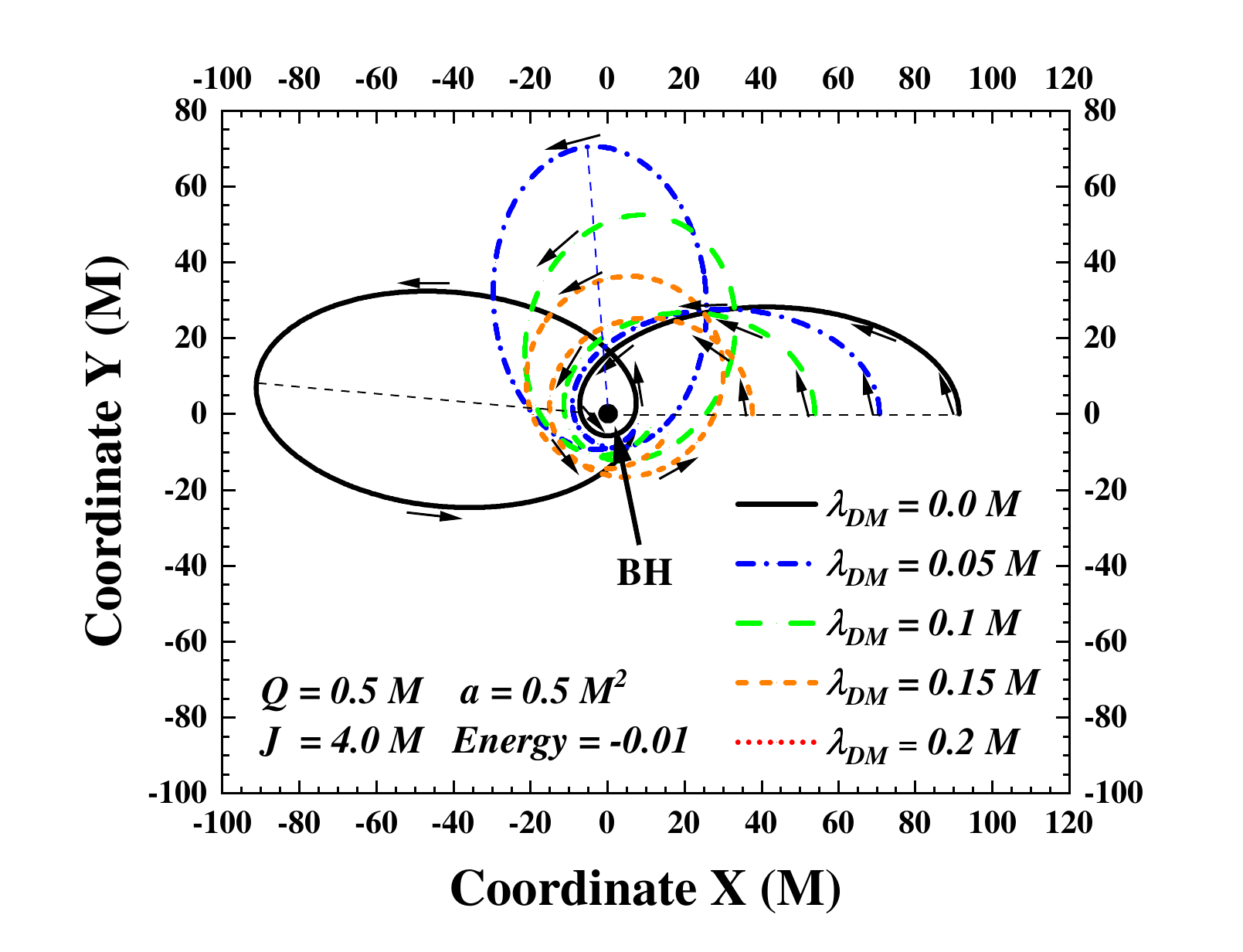}
	\includegraphics[width = 7.5 cm]{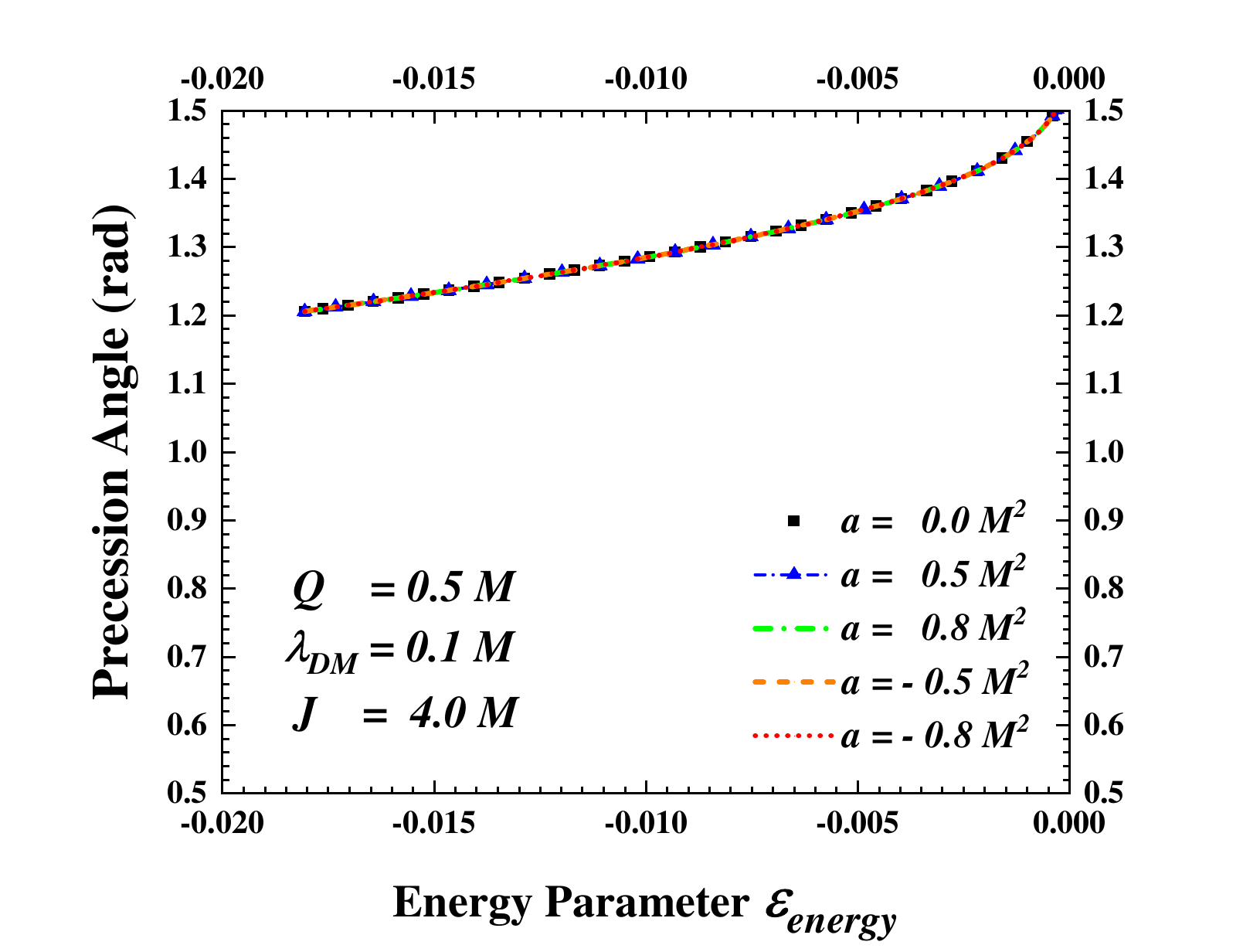}
	\includegraphics[width = 7.5 cm]{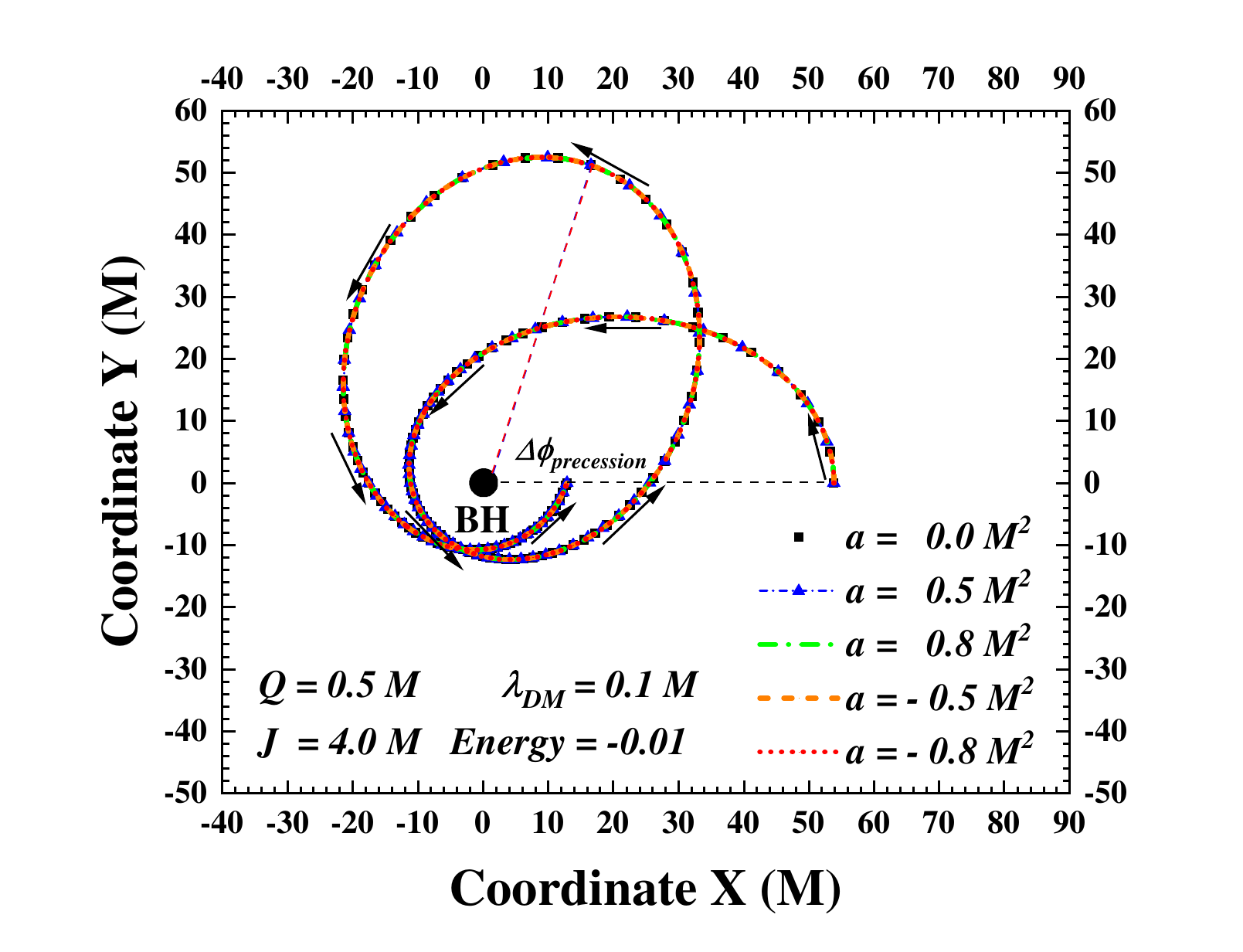}
	\caption{The precession angles and trajectories of massive particle's bound orbits moving around the Euler-Heisenberg black hole in the presence of PFDM. 
	(\textbf{a}) The upper left panel plots the precession angle of orbits affects by black hole charge. (\textbf{b}) The upper right panel illustrates the trajectories of orbits for different black hole charge values. (\textbf{c}) The middle left panel gives the precession angle of orbits affects by dark mater. (\textbf{d}) The middle right panel illustrates the trajectories of orbits for different PFDM parameter values. 
	(\textbf{e}) The lower left panel shows the precession angle of orbits influenced by nonlinear electrodynamics effects / QED effects. (\textbf{f}) The lower right panel illustrates the trajectories of orbits for different nonlinear electrodynamics parameter values. \label{figure precession angle}}
\end{figure}

The figure \ref{figure precession angle} displays numerical results on massive object's bound orbits around the Euler-Heisenberg black hole surround by PFDM, including the precession angles in the left panels and the trajectories of orbits in the right panels.
The left panels of this figure present the variation of precession angle as the changes of total energy parameter $\varepsilon_\text{energy}=\frac{E^{2}-1}{2}$ \footnote{It should be noted that we have substituted $1/2$ in the both side of equation (\ref{reduced differential equation}) such that the energy parameter vanished at infinity, namely $\lim_{r\to \infty}\varepsilon_\text{energy}=0$.}, while the right panels illustrate the trajectories of particle orbits for a given energy parameter $\varepsilon_\text{energy}=-0.01$ \footnote{In the middle right panel of figure \ref{figure precession angle}, the bound orbit does not exist for the $\lambda_{\text{DM}}=0.2M$ case,  because the local minimum of effective potential $V_{\text{min}} < \varepsilon_\text{energy}=-0.01$. So the red dotted curve, which represent the particle orbit for the $\lambda_{\text{DM}}=0.2M$ case, is disappear in the middle right panel. The local minimum of effective potential is given in table \ref{table1}}.
The trajectories of massive object's orbits are obtained by numerically solving the ordinary differential equation (\ref{orbit differential equation}), and the precession angles of orbits are calculated from the integral in expression (\ref{precession angle}).  The upper, middle and lower panels of figure \ref{figure precession angle} highlight the influences from black hole electric charge, dark matter (parameterized by PFDM parameter $\lambda_{\text{DM}}$), and nonlinear electrodynamics effects / QED effects (parameterized by nonlinear electrodynamics parameter $a$) correspondingly. In this figure, we present results on the trajectory of orbits and their precession angles for a selected angular momentum value $J=4M$. The variation trend of precession angles corresponding to other angular momentum values is similar to that for $J=4M$ presented in figure \ref{figure precession angle}.

\begin{table}
	\caption{The local minimum position of effective potential for massive objects moving around Euler-Heisenberg black hole in the presence of PFDM. In this table, the black hole electric charge and nonlinear electrodynamics parameter are chosen to be $Q = 0.5 M$, $a = 0.5 M^{2}$, and conserved angular momentum for massive object's bound orbit is selected as $J=4M$. In this table, we have substituted $1/2$ in the effective potential given in equation (\ref{reduced differential equation}) such that the effective potential for massive object is vanished in the asymptotic flat region (the infinity) $\lim_{r\to \infty}V_\text{eff}(r) = \lim_{r\to \infty} \big\{ \frac{f(r)}{2} \big[ \frac{J^{2}}{r^{2}} + \epsilon \big] - \frac{1}{2} \big\} =0$. \label{table1}}
	\vspace{5pt}
	\begin{tabular}{cccccccc}
		\hline
		PFDM parameter  & $\lambda_{\text{DM}}=0.0M$ & $\lambda_{\text{DM}}=0.05M$ & $\lambda_{\text{DM}}=0.1M$ & $\lambda_{\text{DM}}=0.2M$ 
		\\
		local minimum of $V_{\text{eff}}$ & $V_{\text{min}}=-0.03611$ & $V_{\text{min}}=-0.02539$   & $V_{\text{min}}=-0.01807$  & $V_{\text{min}}=-0.00821$ 
		\\ 
		\hline
	\end{tabular}
\end{table}

The results in figure \ref{figure precession angle} show that both black hole electric charge and PFDM can intensely influence the precession angle of massive object's orbits in the vicinity of Euler-Heisenberg black hole. The upper panels of figure \ref{figure precession angle} suggest that a larger black hole charge reduces the precession angle of orbits. However, unlike the precession angle, the shape of massive object's bound orbits is not violently influenced by black hole electric charge. In the upper right panel of figure \ref{figure precession angle}, the shape and eccentricity of massive object's orbits calculated for different black hole charge values do not exhibit large differences, until the electric charge grows to a value closer the extreme black hole cases (see the $Q = 0.8 M$ case, which corresponds to the dotted line in the upper right panel). On the other hand, the middle panels indicate that the PFDM has great impacts not only on the precession angle of orbits, but also on the shape of massive object's bound orbits. The middle left panel tells that a larger PFDM parameter can significantly reduce the precession angle of bound orbits. The middle right panel shows that the shape, eccentricity and semimajor-axis length of massive object's bound orbits exhibit notable differences when PFDM parameter takes different values. Moreover, in the middle left panel, the proper region for energy parameter $\varepsilon_\text{energy}=\frac{E^{2}-1}{2}$ which allows the existence of bound orbits is also reduced with the increasing of PFDM parameter. This is caused by the local minimum position of effective potential for massive objects. For any massive object's bound orbits with precession, the total energy $\varepsilon_\text{energy}$ in this orbit must exceed the local minimum of effective potential. The local minimum values of effective potential for massive objects moving around Euler-Heisenberg black hole in the presence of PFDM (for a selected angular momentum value $J = 4 M$ and several PFDM parameter values) are listed in table \ref{table1}.
Furthermore, unlike the black hole electric charge and PFDM, the nonlinear electrodynamics effects / QED effects (parameterized by $a$) have very tiny impacts on the trajectories of massive object's orbits and their procession angles, which can be directly observed from the lower panels of figure \ref{figure precession angle}. This trend is similar to the cases of gravitational deflection and time delay, where the nonlinear electrodynamics effects also produce very tiny influences on these gravitational lensing observables.

\subsection{Black Hole Shadow} \label{section4d} 

\begin{figure}
	\centering
	\includegraphics[width = 8.00 cm]{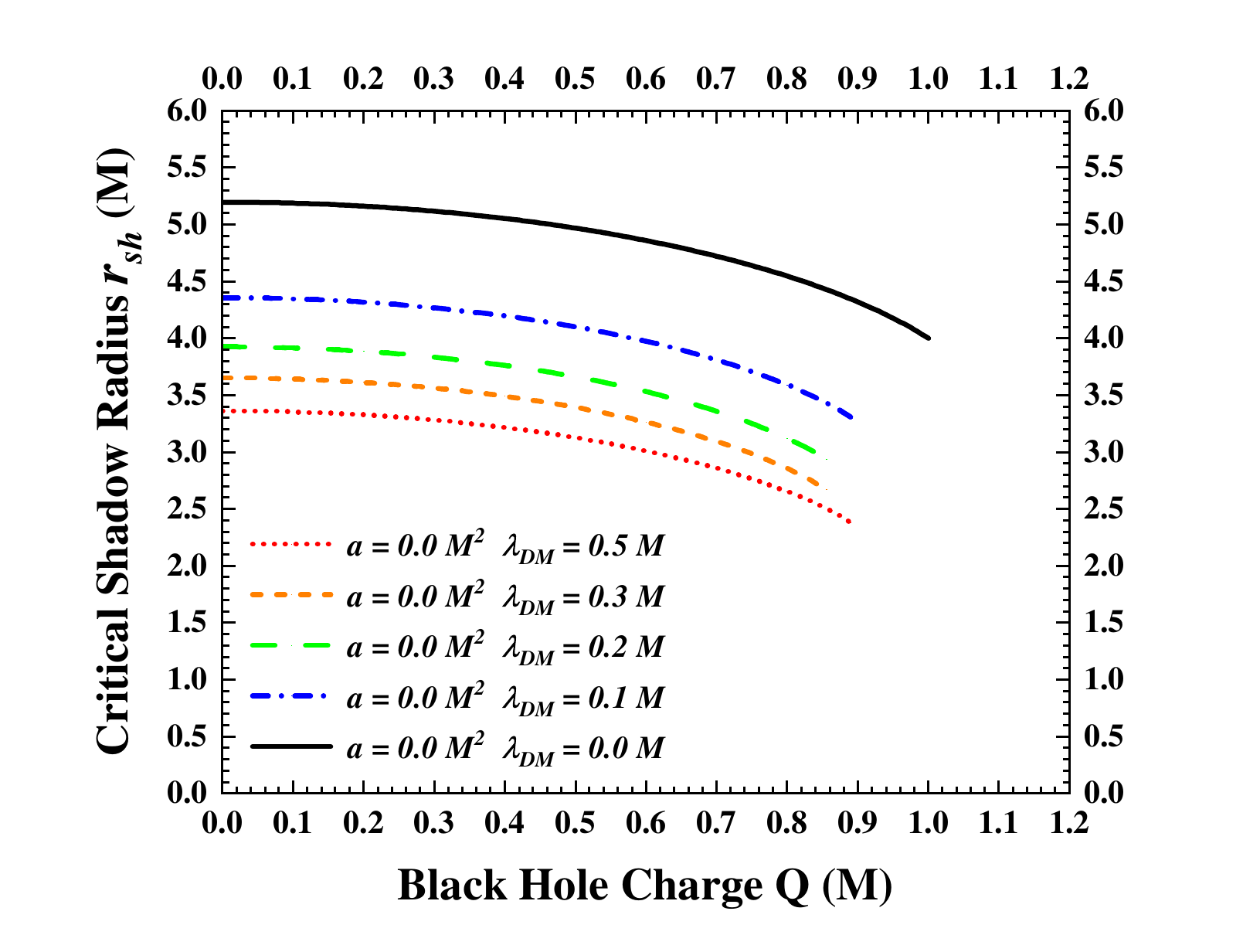}
	\includegraphics[width = 5.75 cm]{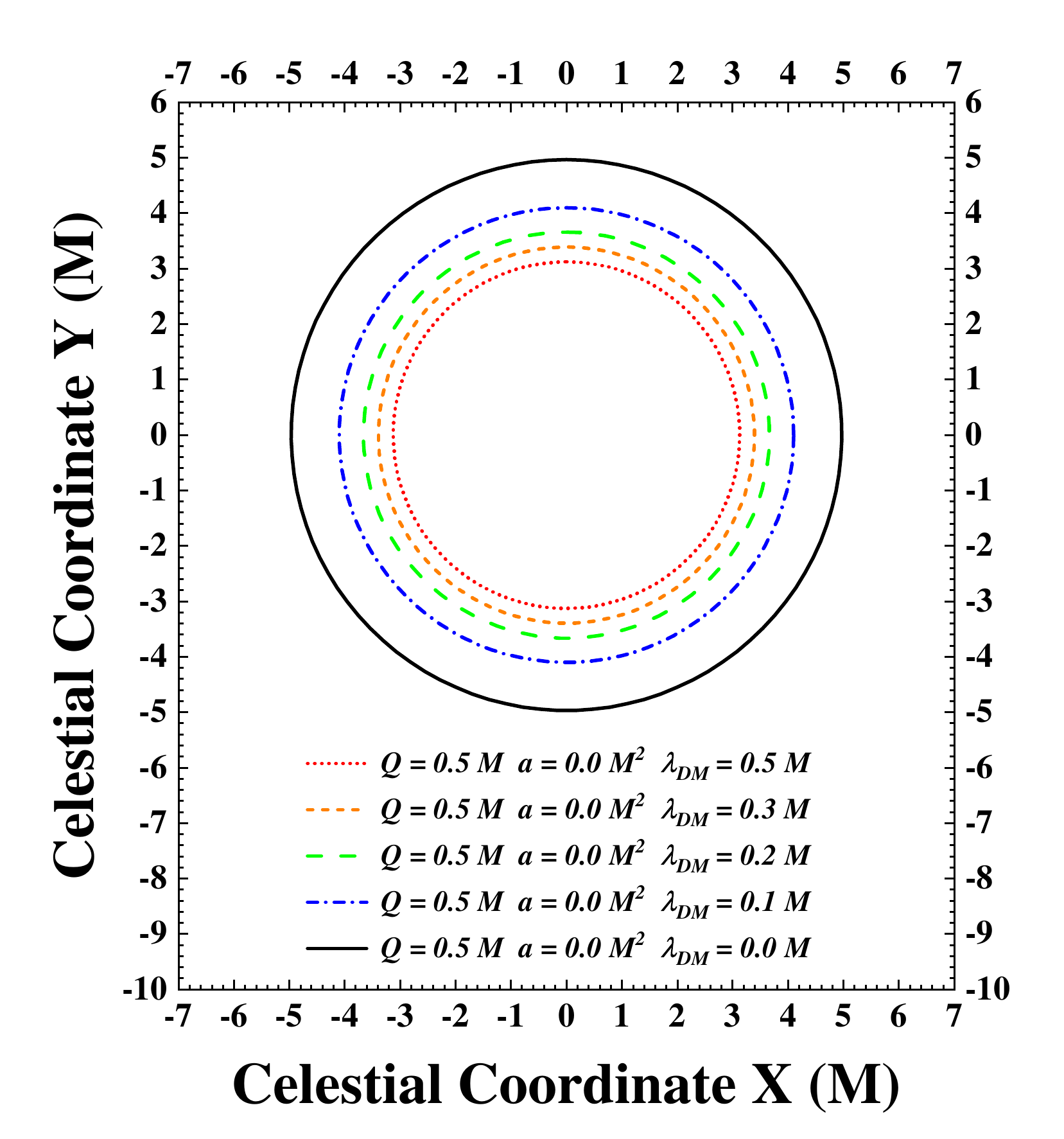}
	\includegraphics[width = 8.00 cm]{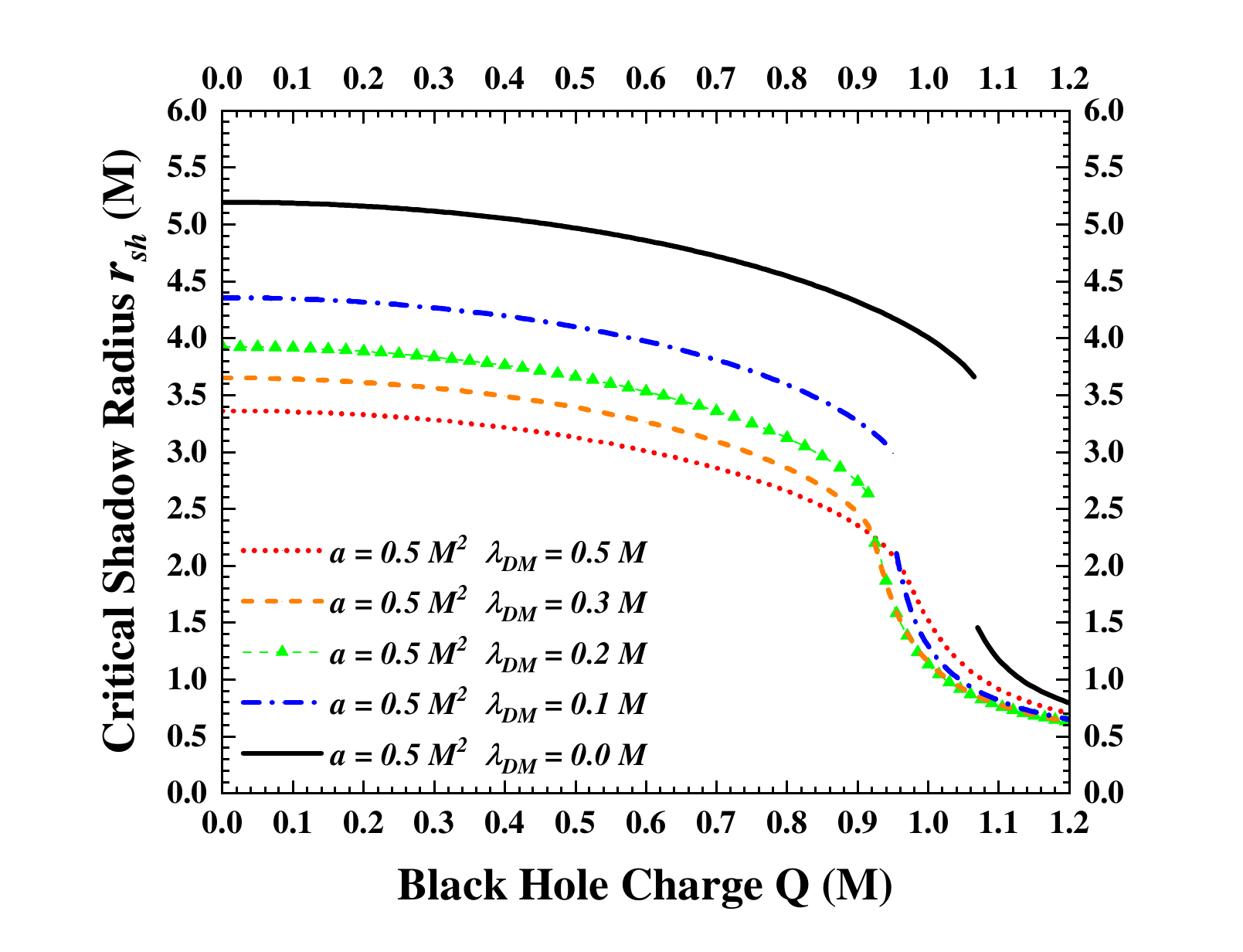}
	\includegraphics[width = 5.75 cm]{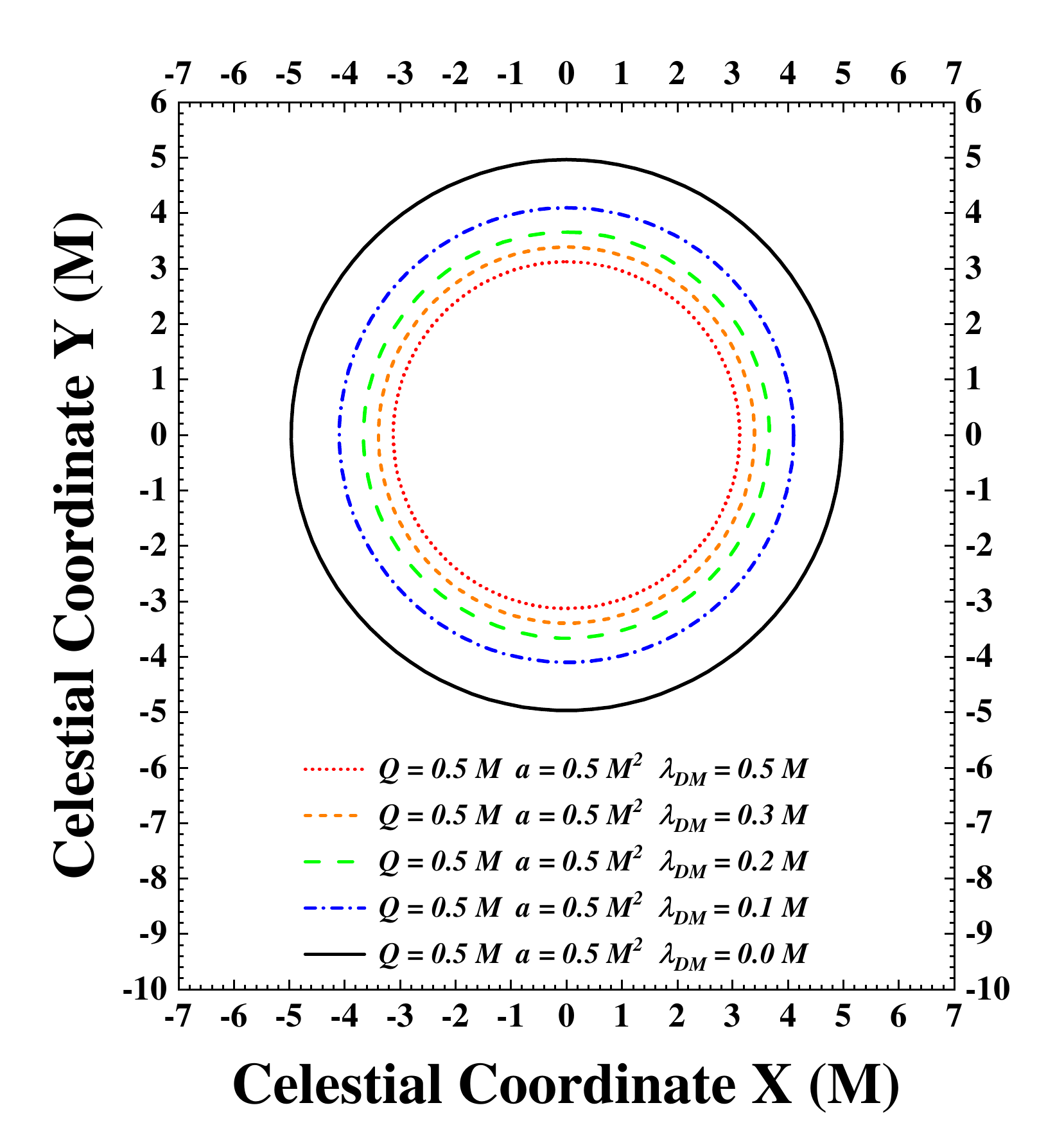}
	\includegraphics[width = 8.00 cm]{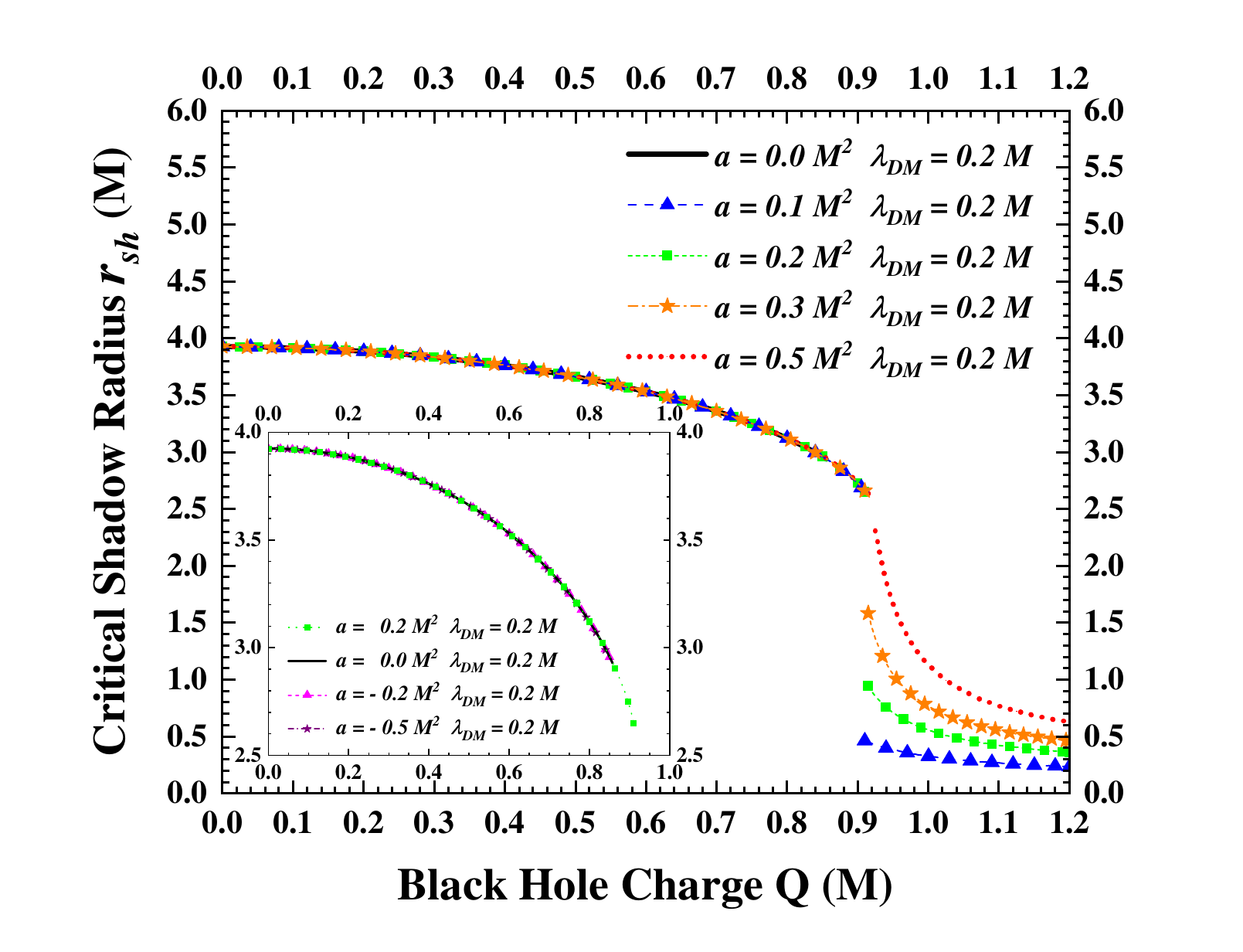}
	\includegraphics[width = 5.75 cm]{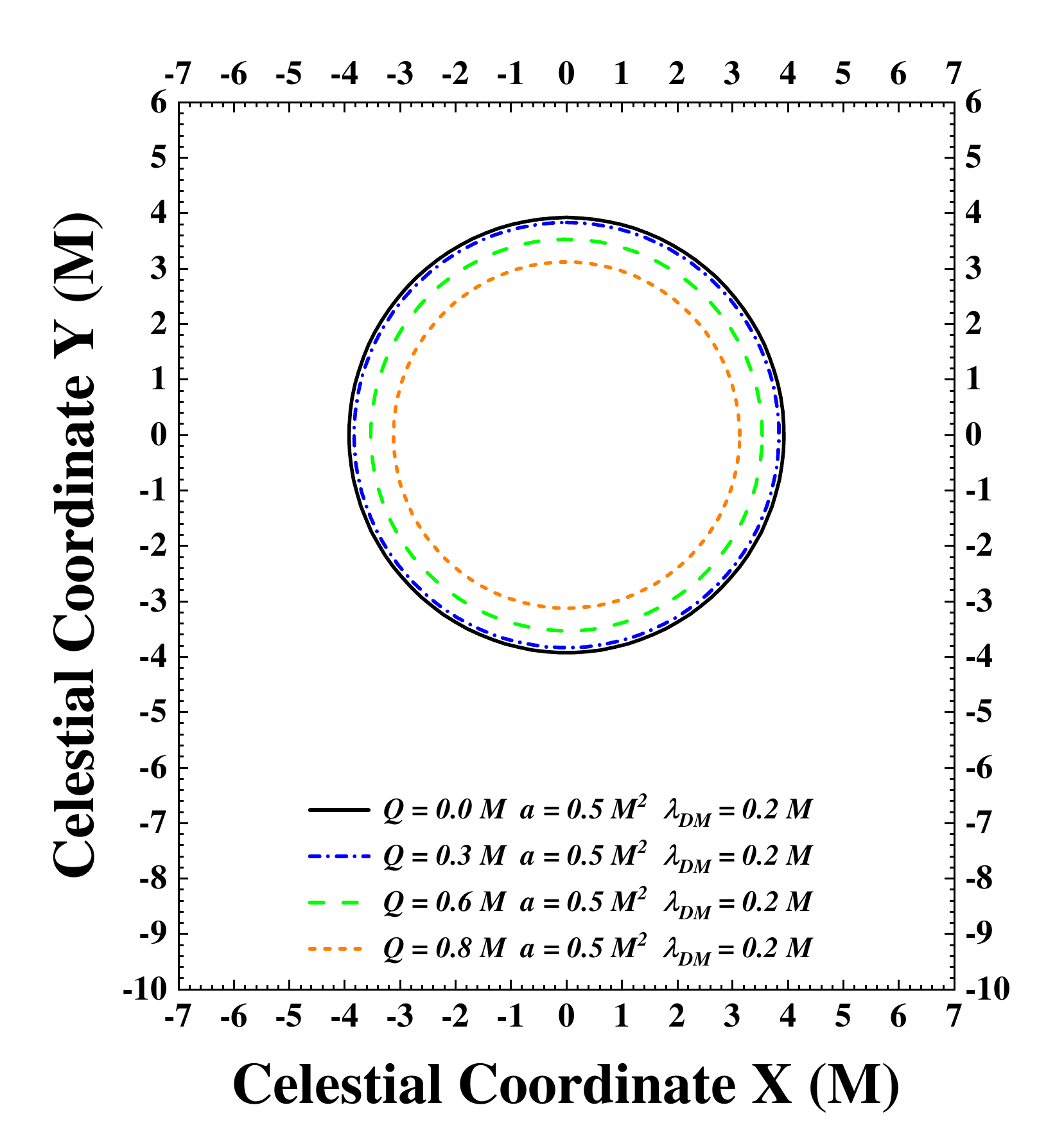}
	\caption{The black hole shadow of Euler-Heisenberg black hole surrounded by PFDM. (\textbf{a}) The upper left panel shows the critical shadow radius affected by PFDM and black hole charge in the absence of nonlinear electrodynamics effects. (\textbf{b}) The upper right panel illustrates the black hole shadow affected by PFDM in the absence of nonlinear electrodynamics effects. (\textbf{c}) The middle left panel shows the critical shadow radius affected by PFDM and black hole charge in the presence of nonlinear electrodynamics effects. (\textbf{d}) The middle right panel illustrates the black hole shadow affected by PFDM in the presence of nonlinear electrodynamics effects. (\textbf{e}) The lower left panel presents the critical shadow radius influenced by black hole charge and nonlinear electrodynamics effects. (\textbf{f}) The lower right panel shows the black hole shadow influenced by black hole electric charge. In all panels, the horizontal and vertical axes are given in unit of black hole mass. \label{figure shadow}}
\end{figure} 

\begin{figure}
	\centering
	\includegraphics[width = 0.495 \textwidth]{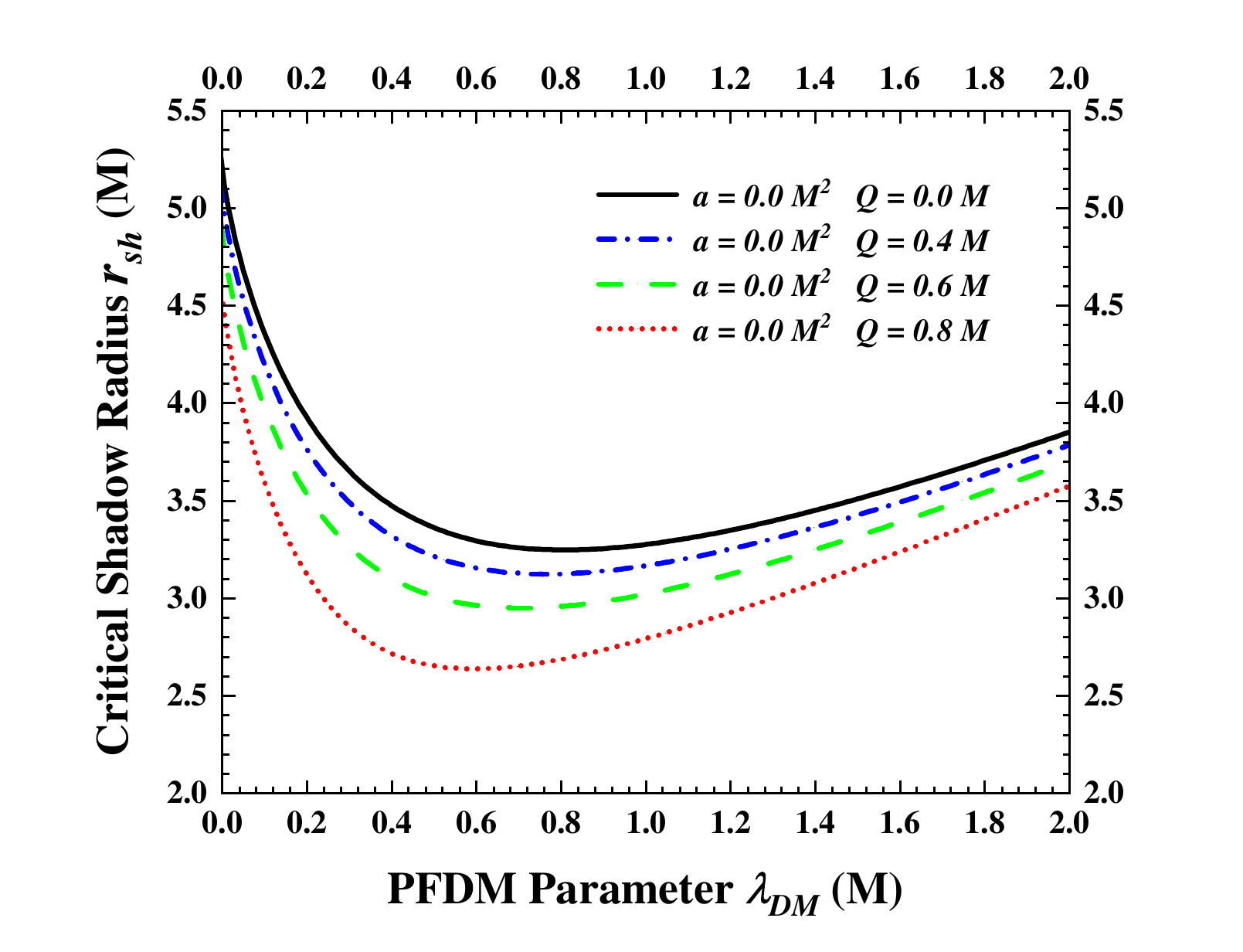}
	\includegraphics[width = 0.495 \textwidth]{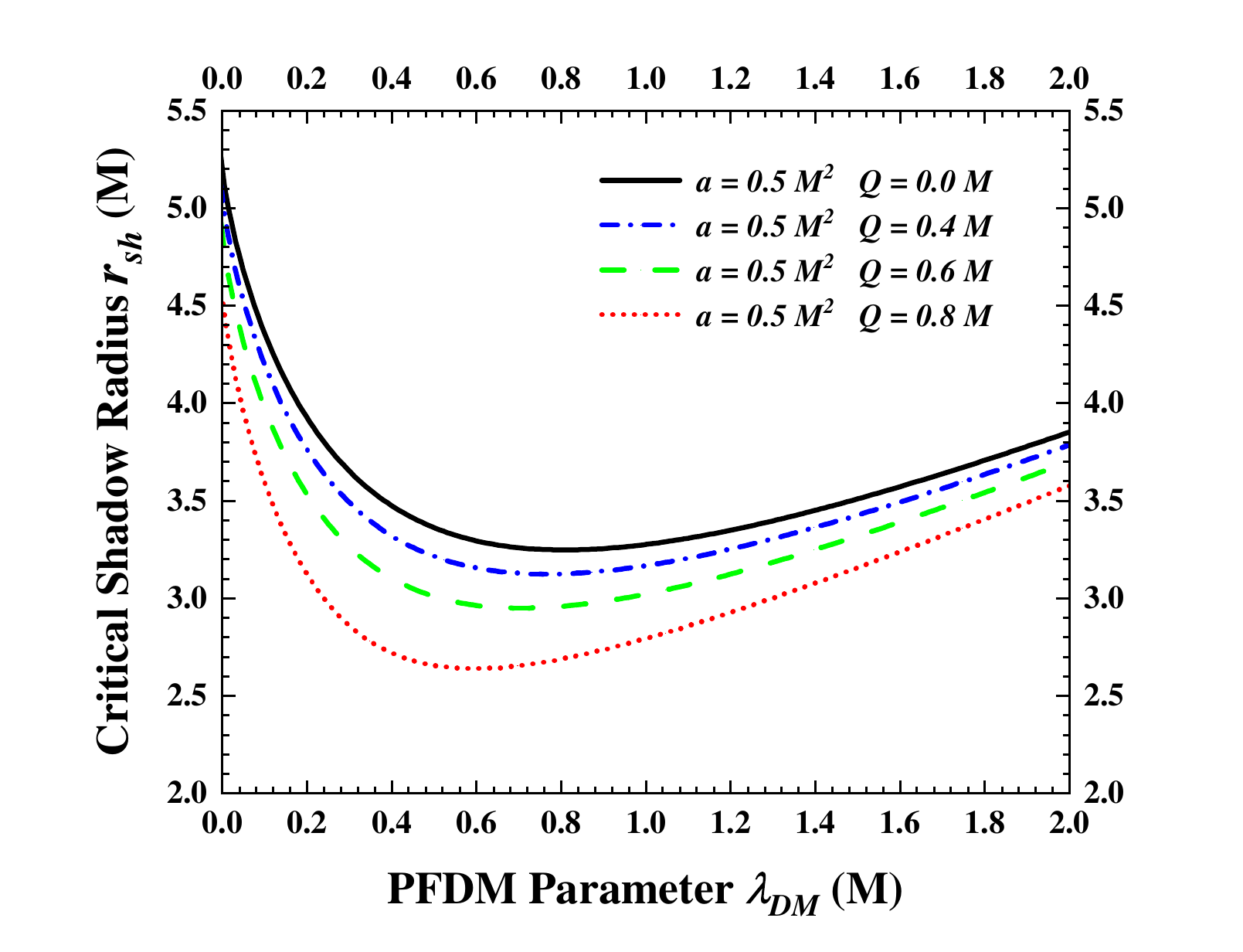}
	\caption{The black hole shadow of Euler-Heisenberg black hole surrounded by PFDM. This figure exhibits the variation of black hole critical shadow radius with respect to PFDM parameter. (\textbf{a}) The left panel shows the critical shadow radius for Euler-Heisenberg black hole surrounded by PFDM in the absence of nonlinear electrodynamics effects (with $a=0$). (\textbf{b}) The right panel shows the critical shadow radius for Euler-Heisenberg black hole surrounded by PFDM in the presence of nonlinear electrodynamics effects (with $a = 0.5 M^{2}$).  \label{figure7}}
\end{figure}

This subsection discusses the black hole shadow for Euler-Heisenberg black hole surrounded by PFDM. The numerical results on black hole shadow for Euler-Heisenberg black hole surrounded by PFDM are presented in figure \ref{figure shadow}. The left panels plot the critical shadow radius $r_{sh}$ calculated using expression (\ref{critical shadow radius}), and the right panels illustrate the black hole shadow in celestial coordinates, highlighting the black hole shadow affected by three black hole parameters (the black hole electric charge $Q$, PFDM parameter $\lambda_{\text{DM}}$ and nonlinear electrodynamics parameter $a$). From figure \ref{figure shadow}, it is reasonable to draw the following conclusions. Firstly, the increasing of black hole electric charge diminishes the apparent size of black hole shadow. Secondly, the lower left panel shows that the nonlinear electrodynamics parameter $a$ has very tiny influences on the black hole shadow size. For the same electric charge and PFDM parameter, the critical shadow radius calculated with different nonlinear electrodynamics parameter values do not exhibit any obvious differences. Thirdly, when PFDM parameter is not vary large, such as $\lambda_{\text{DM}} \le 0.5 M$ in upper and middle panels of figure \ref{figure shadow}, the size of black hole shadow is enlarged as PFDM parameter reduces. However, this statement is no longer valid when PFDM reaches a large critical point, see figure \ref{figure7}. 

To further explore the dark matter effects on black hole shadow size, we use figure \ref{figure7} to show the critical shadow radius changed with respect to PFDM parameter, both in the presence and absence of nonlinear electrodynamics effects. It can be observed that when PFDM parameter is smaller than a critical value ($\lambda_{\text{DM}} < \lambda_{\text{critical}}$), the black hole shadow radius reduces with the increasing of PFDM parameter. However, when PFDM parameter exceeds this critical value ($\lambda_{\text{DM}} > \lambda_{\text{critical}}$), the black hole shadow radius gets magnified as the increasing of PFDM parameter. This seems to be a universal property of the PFDM effects on black hole shadows. Recent studies on charged RN black hole surrounded by PFDM have reported similar conclusions as what we have presented here (the existence of critical point $\lambda_{\text{critical}}$) \cite{Das2022,QiaoCK2023,Das2024}. Furthermore, figure \ref{figure7} shows that the critical value of PFDM parameter is roughly $ 0.5 M < \lambda_{\text{critical}} < 0.8 M$ for different black hole charge values. Since the nonlinear electrodynamics effects have tiny influences on black hole shadow size, the left and right panels of figure \ref{figure7} exhibit very similar results and tendency. 

Interestingly, the black hole shadow results for Euler-Heisenberg black hole surrounded by PFDM in the presence and absence of nonlinear electrodynamics effects / QED effects could also exhibit a slight difference. In the absence of nonlinear electrodynamics effects, because of the emergence of naked singularity when black hole charge $Q > Q_{\text{extreme}}$, the black hole shadow disappears for large black hole electric charge. On the other hand, in the presence of nonlinear electrodynamics effects, the black hole shadow behavior becomes more complicated. When the nonlinear electrodynamics parameter takes a negative value ($a<0$), the Euler-Heisenberg black hole (surrounded by PFDM) still possesses naked singularity for a large black hole charge. In such cases, the shadow for Euler-Heisenberg black hole disappears when black hole charge exceeds its extreme value $Q>Q_{\text{extreme}}$ (with the extreme charge a little bit larger than the RN black hole case), which is shown in the lower left panel of figure \ref{figure shadow}. However, when the nonlinear electrodynamics parameter takes a positive value ($a>0$), the Euler-Heisenberg black hole (surrounded by PFDM) does not produce any naked singularities (see Appendix \ref{appendix1} for the discussions on horizons), the black hole shadow calculated from expression (\ref{critical shadow radius}) can still appear for a larger black hole charge, but the critical shadow radius may undergo a discontinuity under some circumstances. In the middle left panel of figure \ref{figure shadow}, when PFDM parameter takes the values $\lambda_{\text{DM}} = 0.0M, 0.1M, 0.2M$, the critical shadow radius calculated using expression (\ref{critical shadow radius}) is not continuous when black hole charge reaches a critical value $Q=Q_{\text{discontinuity}}$. This is caused by the discontinuity of unstable photon sphere positions in the vicinity of Euler-Heisenberg black hole. When black hole charge exceeds this critical value, the number of unstable photon spheres around Euler-Heisenberg black hole undergoes a sudden change 
\footnote{When black hole charge satisfies $Q < Q_{\text{discontinuity}}$, there are two unstable photon spheres near the Euler-Heisenberg black hole (surrounded by PFDM), and the outer branch of unstable photon sphere determines the black hole shadow size via expression (\ref{critical shadow radius}). When black hole charge becomes $Q > Q_{\text{discontinuity}}$, there is only one unstable photon sphere. The outer branch of unstable photon spheres in the $Q < Q_{\text{discontinuity}}$ cases is disappeared, which makes the unstable photon sphere position undergo a sudden change when black hole charge passes the critical point $Q=Q_{\text{discontinuity}}$.}. 
Once the black hole charge comes across the discontinuity point (namely $Q > Q_{\text{discontinuity}}$), it is questionable that the results obtained using expression (\ref{critical shadow radius}) could still represent a conventional ``black hole shadow'' or not. This issue deserves a more deeply study in the future.

\section{Conclusions and Perspectives} \label{section5}

In this work, the gravitational lensing of Euler-Heisenberg black hole surrounded by PFDM is studied. This kind of black hole solution, which is solved from the gravitational field equation coupled with nonlinear electromagnetic field and dark matter field, could give rise to the nonlinear electrodynamics effects and PFDM effects in charge black hole systems. It provides us a simple way to investigate the interplay of nonlinear electrodynamics effects (or QED effects) and dark matter effects on the studies of black holes. These effects coming from nonlinear electrodynamics and dark matter can be parameterized by parameters $a$ and $\lambda_{\text{DM}}$ respectively. Particularly, the nonlinear electrodynamics effects have significant impact on the horizon structure of charged black hole. In the presented work, we mainly focus on the gravitational lensing observables for Euler-Heisenberg black hole surrounded by PFDM. The gravitational deflection angle of light, time delay of light, precession angle and trajectories of massive object's bound orbits, black hole shadow are calculated numerically from geodesics.

Our numerical results presented in this work show that influences coming from black hole electric charge, nonlinear electromagnetic effects, dark matter effects exhibit a similar tendency for gravitational lensing observables (such as gravitational deflection angle of light and time delay of light), precession angle of massive object's bound orbits, and the black hole shadow size. Firstly, the Euler-Heisenberg black hole surround by PFDM with a larger electric charge could reduce the gravitational deflection angle, time delay of light, precession angle of bound orbits and black hole shadow radius. Secondly, the nonlinear electrodynamics effects parameterized by $a$ has very tiny influences on these quantities. 
Thirdly, the most significant impacts on these observables are coming from the dark matter effects. The gravitational deflection angle, time delay of light, precession angle of bound orbits and black hole shadow radius are all greatly changed when PFDM parameter $\lambda_{\text{DM}}$ is varied. Particularly, a larger PFDM parameter could greatly reduce the gravitational deflection angle, time delay of light and precession angle of bound orbits. This can be explained from the spacetime metric function (\ref{spacetime metric component Euler-Heisenberg}) for Euler-Heisenberg black hole surrounded by PFDM, in which the PFDM parameter plays a role of ``effective mass''. A positive PFDM parameter contributes to a negative ``effective mass'' in the spacetime metric, therefore it can diminish the gravitational lensing observables (such as gravitational deflection angle and time delay of light). 
Moreover, the dark matter also has notable influences on the shape and eccentricity of massive object's bound orbits. For the black hole shadow size, when PFDM parameter is smaller than a critical value, the black hole shadow radius gets reduced as the increasing of PFDM parameter. When PFDM parameter exceeds this critical value, the black hole shadow radius could be magnified when PFDM parameter increases. The critical value of PFDM parameter is roughly $ 0.5 M < \lambda_{\text{critical}} < 0.8 M$ for different black hole charge values.

We hope that the results in our present work can deepen our understanding of the nonlinear electrodynamics effects / QED effects and dark matter effects on the gravitational lensing of charged black hole systems. Hopefully, it may have potential applications for theoretical and observational probing for dark matter effects in black holes and gravitation. Since the dark matter could significantly influence the gravitational lensing of supermassive black holes, the PFDM parameter can be constrained from gravitational lensing observations, by analyzing and comparing the observed data with theoretical predictions, especially with the help of accumulating observational data for supermassive black hole in galaxy centers. However, the nonlinear electrodynamics effects have very tiny influences on gravitational lensing observables, so it would be not easy to probe the nonlinear electrodynamics in gravitational lensing observations (unless very high-precision observational data are employed).

Furthermore, there are higher order QED corrections go beyond Euler-Heisenberg theory, which may contribute to the $1/r^{b}$ (with $b>6$) term in the spacetime metric of charged black hole systems. Black hole solutions with higher order QED contributions beyond the Euler-Heisenberg effective theory would be deduced and solved from a fundamental quantum gravity theory, and those black hole solutions may be more complicated than the Euler-Heisenberg black hole presented in this work. The seeking of such black hole solutions with the higher order QED corrections deserves future studies, which could provide us insights into the quantum effects of black hole systems and enhance our understanding of gravitational and electromagnetic interactions.

\section*{Acknowledgment}

The authors thank Song-Lin Lyu and Guang-Zhou Guo for helpful discussion and comments on the numerical schemes. This research was funded by the Natural Science Foundation of Chongqing Municipality (Grant No. CSTB2022NSCQ-MSX0932), the Scientific and Technological Research Program of Chongqing Municipal Education Commission (Grant No. KJQN202201126 and No. KJQN202301164), the Scientific Research Program of Chongqing Municipal Science and Technology Bureau (the Chongqing “zhitongche” program for doctors, Grant No. CSTB2022BSXM-JCX0100), the Scientific Research Foundation of Chongqing University of Technology (Grant No. 2020ZDZ027) and the Research and Innovation Team Cultivation Program of Chongqing University of Technology (Grant No. 2023TDZ007).

\appendix

\section[\appendixname~\thesection]{Horizons for Euler-Heisenberg Black Hole Surrounded by Perfect Fluid Dark Matter \label{appendix1}}

\begin{figure}[t]
	\centering
	\includegraphics[width = 0.495 \textwidth]{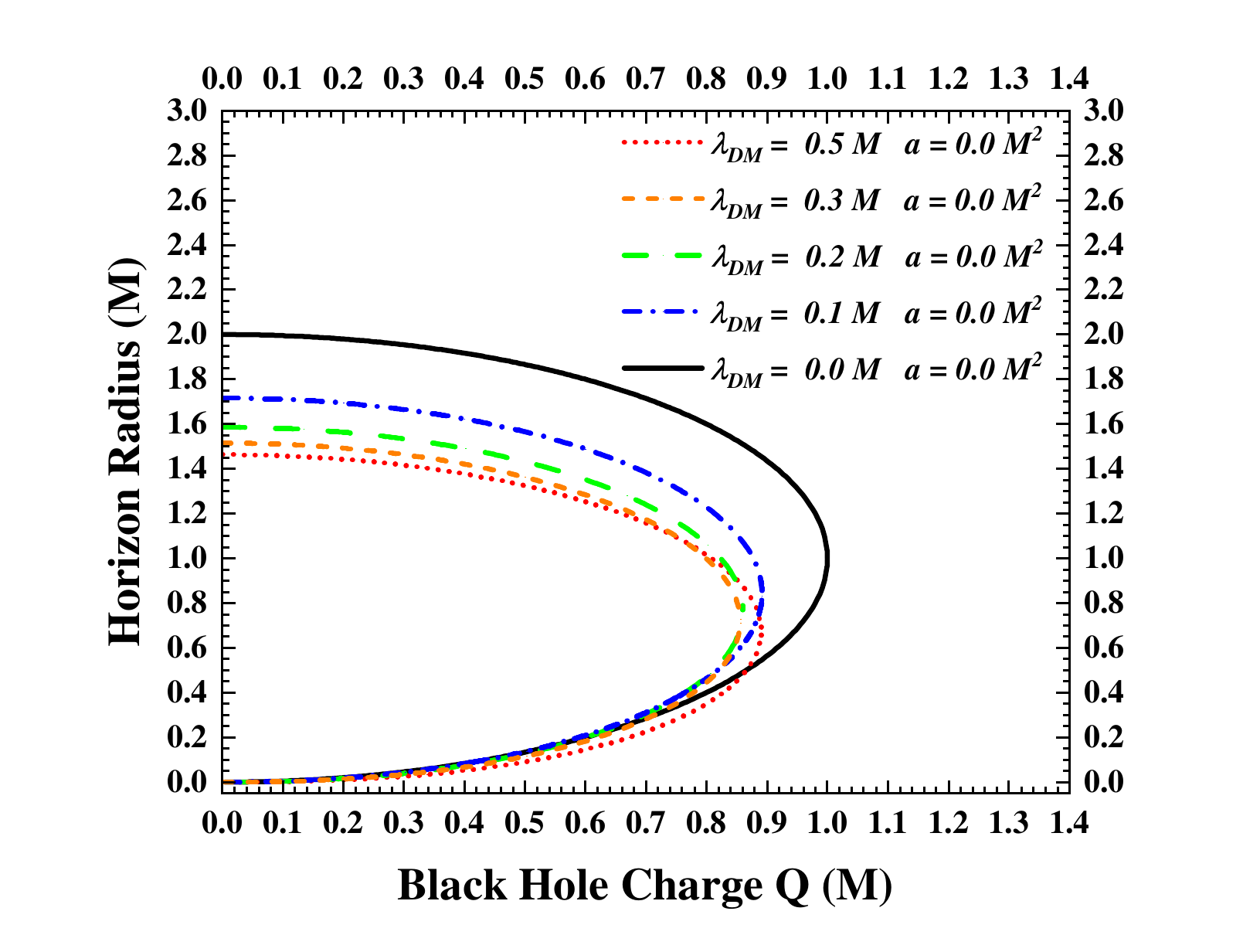}
	\includegraphics[width = 0.495 \textwidth]{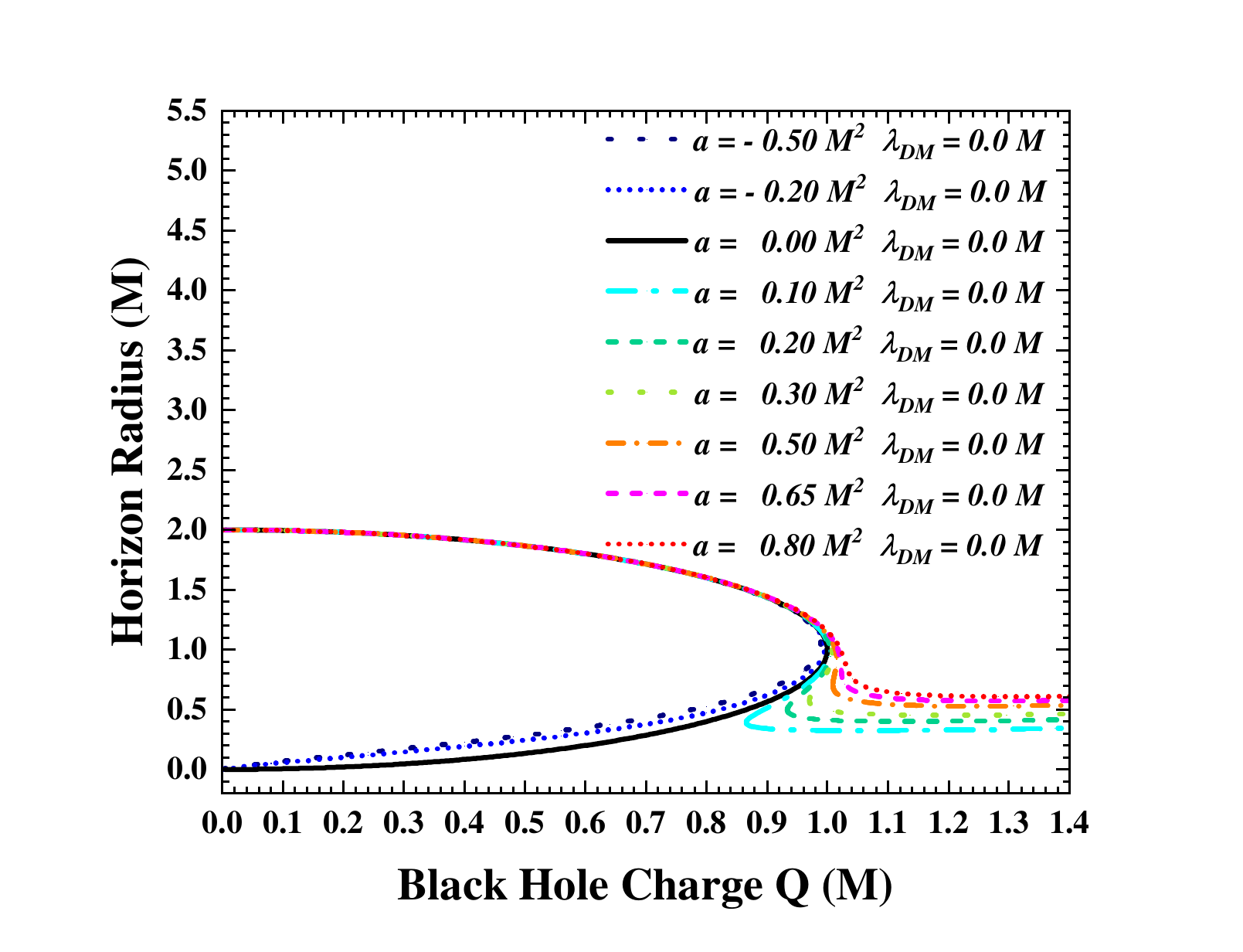}
	\includegraphics[width = 0.495 \textwidth]{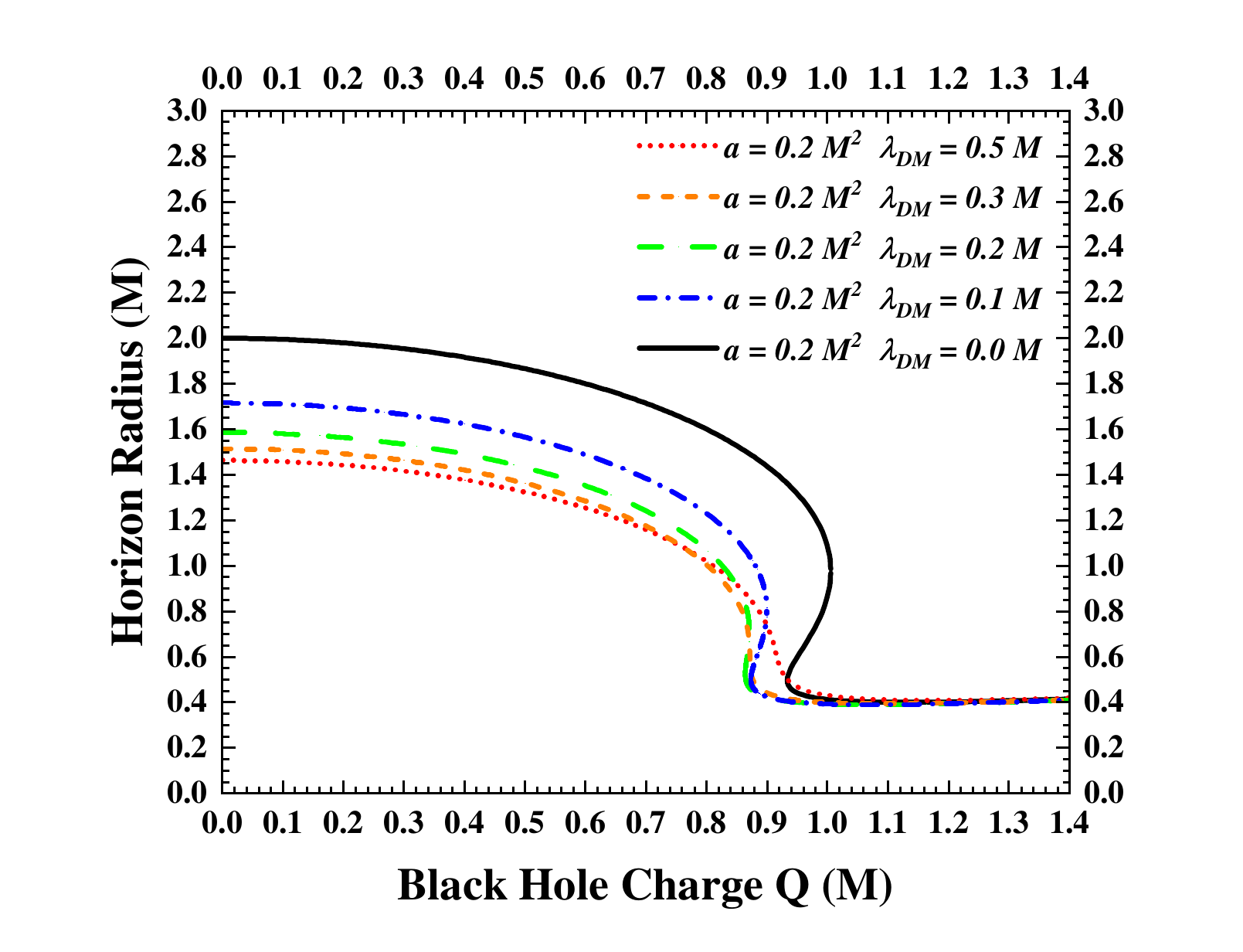}
	\includegraphics[width = 0.495 \textwidth]{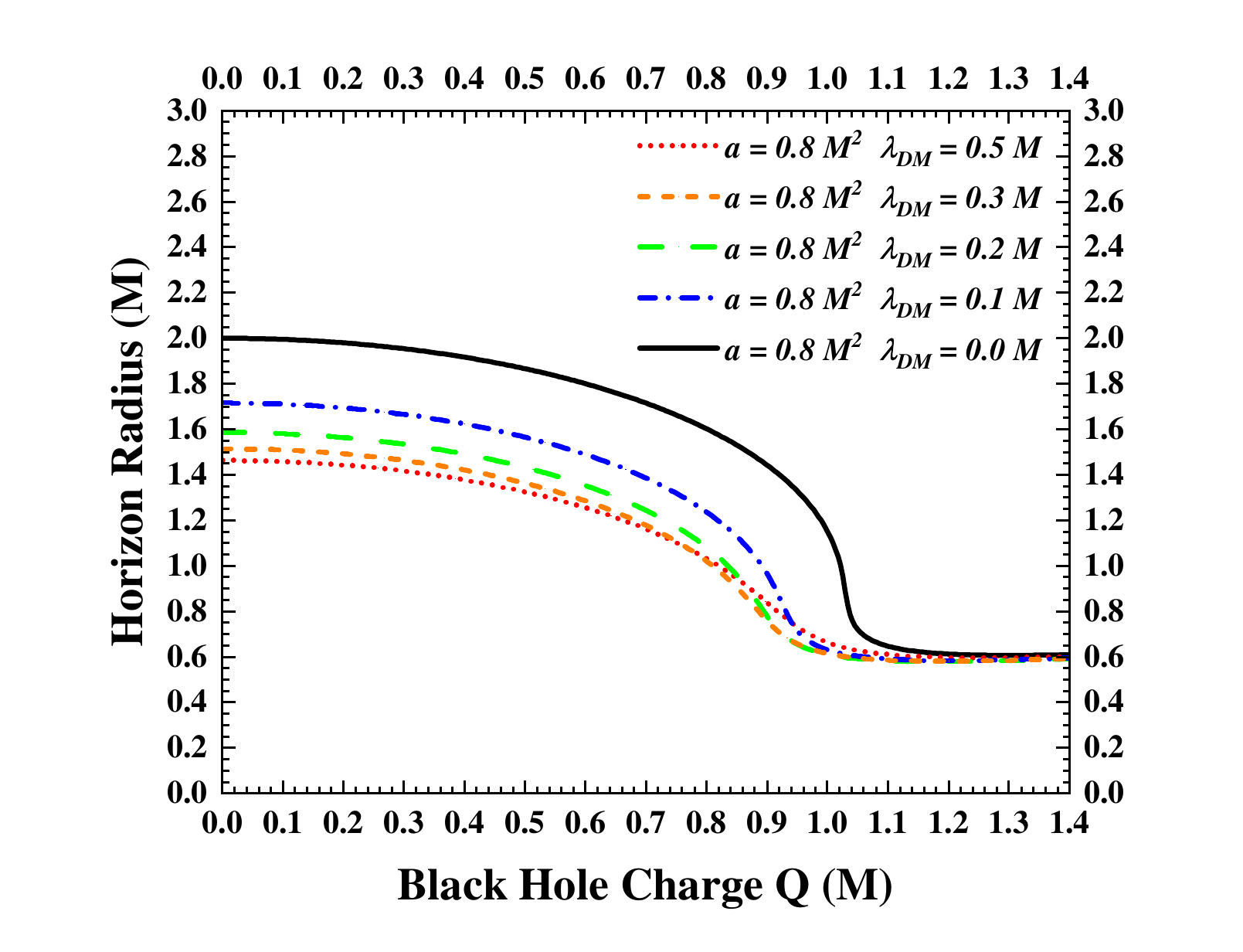}
	\caption{The horizons of Euler-Heisenberg black hole surrounded by PFDM. (\textbf{a}) The upper left panel shows the charged black hole's horizons affected by PFDM parameter $\lambda_{\text{DM}}$ without the nonlinear electrodynamics effects. (\textbf{b}) The upper right panel shows the Euler-Heisenberg black hole's horizons affected by nonlinear electrodynamics effects without PFDM. (\textbf{c-d}) The lower left and lower right panels plot the horizon radius of Euler-Heisenberg black hole in the presence of dark matter and nonlinear electrodynamics (with $a=0.2M^{2}$ and $a=0.8M^{2}$ respectively). In all panels, the horizontal and vertical axes are given in unit of black hole mass. \label{figure A1}}
\end{figure}

This appendix presents results on horizons for Euler-Heisenberg black hole surrounded by PFDM. The radii of horizons can be solved from the metric component function 
\begin{equation}
	f(r_{H}) = 1 - \frac{2M}{r_{H}} + \frac{Q^2}{r_{H}^2} - \frac{aQ^4}{20r_{H}^6}
	+ \frac{\lambda_{\text{DM}}}{r_{H}} \cdot \ln \frac{r_{H}}{|\lambda_{\text{DM}}|} = 0 .
	\label{spacetime horizon}
\end{equation}
In the absence of nonlinear electrodynamics effects / QED effects and dark matter effects (with $a \neq 0$ and $\lambda_{\text{DM}}=0$), the spacetime reduces to a conventional RN spacetime, resulting in two horizons for $Q<M$, a single horizon for the extreme case $Q=M$, and a naked singularity at the center for $Q>M$. However, in the presence of nonlinear electrodynamics effects and dark matter effects, the horizon structure can be strongly influenced. In figure \ref{figure A1}, we plot the horizon radius for Euler-Heisenberg black hole surrounded by PFDM, which is calculated by numerically solving equation (\ref{spacetime horizon}). The multiple panels of this figure highlight the influences coming from dark matter effects and nonlinear electrodynamics effects.

In the absence of nonlinear electrodynamics effects (with $a=0$), the dark matter effects do not have obvious influences on the structure of horizons, which can be easily observed in the upper left panel of figure \ref{figure A1}. Without the nonlinear electrodynamics effects, charged black hole surrounded by PFDM still possesses two horizons (inner and outer horizons) when $0 < Q < Q_{\text{extreme}}$. When the black hole charge exceeds the extremal black hole charge ($Q > Q_{\text{extreme}}$), the spacetime becomes horizonless and produces a naked singularity in the center $r=0$. This is similar to the naked singularity in conventional RN spacetime. In the presence of PFDM, the extremal black hole charge $Q_{\text{extreme}}$ becomes less than the conventional RN case. This can be explained from the ``effective mass'' role that PFDM played in spacetime metric (\ref{spacetime metric component Euler-Heisenberg}), as we have explained in section \ref{section4}. A positive PFDM parameter contributes to a negative ``effective mass'' in spacetime metric, and the corresponding extremal black hole charge is less than black hole mass ($Q_{\text{extreme}} < M$) for a given positive PFDM parameter ($\lambda_{\text{DM}} > 0$).  

In the presence nonlinear electrodynamics effects (with $a \neq 0$), the horizon structure for charged Euler-Heisenberg black holes can be dramatically changed, which is clearly shown in the upper right panel of figure \ref{figure A1}. When the nonlinear electrodynamics parameter take a negative value $a<0$, the horizons for Euler-Heisenberg black hole are similar to the RN black hole, possessing inner and outer horizons for black hole charge $Q<Q_{\text{extreme}}$ and giving rise to naked singularity when $Q>Q_{\text{extreme}}$. However, for a positive nonlinear electrodynamics parameter ($a>0$), the horizons for charged black hole have three main differences compared to the RN black hole. Firstly, the naked singularity no longer exists when nonlinear electrodynamics parameter takes a positive value ($a>0$). In such cases, there is at least one horizon in Euler-Heisenberg black hole for any given black hole charge. Secondly, for a small black hole charge, the Euler-Heisenberg black hole has only one horizon, unlike the two horizon behavior (the inner and outer horizons) for pure RN spacetime with $a=0$. In the upper right panel of figure \ref{figure A1}, various curves correspond to positive nonlinear electrodynamics parameter values ($a>0$) in the branch $M < r_{H} < 2M$ and $Q < M$ are almost overlapped with each other, but the RN black hole with $a=0$ has a second branch at $0 < r_{H} < M$ and $Q < M$ (labeled by black solid curve in the upper right panel). Thirdly, if the black hole electric charge falls into the region $0.85M < Q < 1.02M$, the Euler-Heisenberg black hole with a small positive nonlinear electrodynamics parameter ($0 < a < 0.5M^{2}$) could have three horizons. However, a larger nonlinear electrodynamics parameter (for instance, the $a = 0.65M^{2}$ and $a = 0.8M^{2}$ in the upper right panel of figure \ref{figure A1}) could prevent the occurrence of three horizons in Euler-Heisenberg black hole, producing a single horizon for arbitrary electric charge values.

When taking into account both nonlinear electrodynamics effects (with a positive nonlinear electrodynamics parameter $a>0$) and dark matter effects (with $\lambda_{\text{DM}} > 0$), the horizons of Euler-Heisenberg black hole surrounded by PFDM are plotted in the lower panels of figure \ref{figure A1}, for nonlinear electrodynamics parameters $a=0.2M^{2}$ and $a=0.8M^{2}$ respectively \footnote{When the nonlinear electrodynamics parameter becomes negative, the horizon behavior for Euler-Heisenberg black hole surrounded by PFDM is similar to the RN black hole surrounded by PFDM given in the upper left panel of figure \ref{figure A1}}. Comparing the positions of horizon for Euler-Heisenberg black hole without dark matter (in the upper right panel) and those in the presence of dark matter (in the lower left and lower right panels), we can obtain the following conclusions. In the presence of dark matter and nonlinear electrodynamics, the naked singularity is no longer emerged for arbitrary black hole electric charge, which is the same as the results for Euler-Heisenberg black hole without PFDM. The dark matter influences the horizon size and changes the width of three-horizon region in the parameter space of black hole charge. A positive PFDM parameter ($\lambda_{\text{DM}} > 0$) results in a smaller horizon radius, which can be explained using the ``effective mass'' role of the PFDM played in spacetime metric. Furthermore, a positive PFDM parameter narrows the three-horizon region in the parameter space of black hole charge. For instance, when nonlinear electrodynamics parameter gets the value $a=0.2M^{2}$, the three-horizon region becomes narrowed when $0 < \lambda_{\text{DM}} \leq 0.3 M$ (compared with those without PFDM plotted in the upper right panel), while the $\lambda_{\text{DM}} = 0.5M$ case does not possess a three-horizon region. For a large nonlinear electrodynamics parameter (such as $a=0.8M^{2}$ in the lower right panel), the Euler-Heisenberg black hole in the presence of dark matter does not possess a three-horizon region, no matter what the PFDM parameter is varied.

\section*{Abbreviations}

The following abbreviations are used in this manuscript:

\ \ 

\noindent 
\begin{tabular}{@{}ll}
	QED   & Quantum Electrodynamics              \\
	PFDM  & Perfect Fluid Dark Matter            \\
	RN    & Reissner-Nordstr\"om                 \\
	WIMPs & Weakly Interacting Massive Particles \\
	EHT   & Event Horizon Telescope              \\
\end{tabular}

\end{document}